\documentclass[a4paper,11pt]{article}
\pdfoutput=1 

\usepackage{jheppub} 

\usepackage{hyperref}
\usepackage{graphics}
\usepackage{slashed}
\usepackage{color}
\usepackage[makeroom]{cancel}
\begin{document}

\leftmargin -2cm
\def\choosen{\atopwithdelims..}

\boldmath
\title{Towards stability of NLO corrections \\ in High-Energy Factorization via \\ Modified Multi-Regge Kinematics approximation} \unboldmath

\author{Maxim Nefedov} \emailAdd{maxim.nefedov@desy.de}

\affiliation{{I}{I.} Institut f\"ur Theoretische Physik, Universit\"at Hamburg, Luruper Chaussee 149, 22761 Hamburg, Germany; \\ Samara National Research University, Moskovskoe Shosse,
34, 443086, Samara, Russia}

\abstract{
  The perturbatively-stable scheme of Next-to-Leading order (NLO) calculations of cross-sections for multi-scale hard-processes in DIS-like kinematics is developed in the framework of High-Energy Factorization. The evolution equation for unintegrated PDF, which resums $\log 1/z$-corrections to the coefficient function in the Leading Logarithmic approximation together with a certain subset of Next-to-Leading Logarithmic and Next-to-Leading Power corrections, necessary for the perturbative stability of the formalism, is formulated and solved in the Doubly-Logarithmic approximation. An example of DIS-like process, induced by the operator ${\rm tr}\left[G_{\mu\nu}G^{\mu\nu}\right]$, which is sensitive to gluon PDF already in the LO, is studied. Moderate ($O(20\%)$) NLO corrections to the inclusive structure function are found at small $x_B<10^{-4}$, while for the $p_T$-spectrum of a leading jet in the considered process, NLO corrections are small ($<O(20\%)$) and LO of $k_T$-factorization is a good approximation. The approach can be straightforwardly extended to the case of multi-scale hard processes in $pp$-collisions at high energies.     
         }



\maketitle

\section{Introduction}
\label{sec:Int} 

The High-Energy or $k_T$-factorization formalism, first introduced in~\cite{Gribov:1984tu}, later had been developed~\cite{Collins:1991ty, Catani:1994sq} as a tool to resum higher-order corrections to coefficient functions of Collinear Parton Model, enhanced by large logarithms $\log(1/z)$ of light-cone momentum fraction $z$, of a parton entering into a hard subprocess, relative to the characteristic light-cone momentum component of a final-state of interest. This kind of corrections become more and more important with increasing collision energy, since more phase-space for additional semi-hard emissions opens up. These emissions generate a transverse momentum recoil, which greatly affects kinematic distributions of the final-state of interest, e.g. di- or multi-jet system~\cite{Nefedov:dijet, Bury:forward-dijet, Kutak:4-jet}, pair of heavy-flavoured mesons~\cite{Karpishkov:BB,Maciula:DD} or heavy quarkonia~\cite{He:di-Jpsi}. Therefore, $k_T$-factorization calculation serves as an interesting alternative to fixed-order calculations of such observables in Collinear Parton Model (CPM) or with conventional Parton Showers (PS) (see Ref.~\cite{Buckley:2011ms} for the review). As one can see, e.g. from references cited above, the $k_T$-factorization canculation with a judicious choice of unintegrated-Parton Distribution Function (UPDF) quite often leads to a good description of various correlation observables already in the leading order (LO) in $\alpha_s$, as opposed to situation in CPM, where e.g. the description of $\Delta\phi$-spectrum can be quite poor even at Next-to-Leading order (NLO) and significant initial-state PS effects have to be taken into account.         

  However a multitude of phenomenological approaches to determine the UPDF (see Ref.~\cite{Hautmann:TMDlib} for a comprehensive list), their seeming inconsistency with each-other and lack of a practical formalism to perform Next-to-Leading order calculations, which goes beyond the results of pioneering papers~\cite{Ostrovsky:1999kj, Bartels:2006hg}, always where major drawbacks of High-Energy factorization program. The perturbative instability of Balitsky-Fadin-Kuraev-Lipatov(BFKL)-formalism~\cite{BFKL1, BFKL2, BFKL3}, first observed in a celebrated calculation of NLO BFKL kernel~\cite{NLO-BFKL, NLOCiafaloni1, NLOCiafaloni2} is a main reason of a slow development of $k_T$-factorization beyond LO. The main source of large NLO corrections to the BFKL kernel was immediately identified in the Ref.~\cite{Salam98}, these are large logarithms of transverse momentum, coming from the collinear region of the NLO correction, which are not reproduced by the iteration of LO kernel. Resummation of this large logarithms requires development of an approach unifying BFKL and Dokshitzer-Gribov-Lipatov-Altarelli-Parizi(DGLAP)~\cite{DGLAP1,DGLAP2,DGLAP3} dynamics, which is a task of formidable complexity. Several approaches to this problem had been proposed~\cite{Brodsky:1998kn, Altarelli:1999vw, RGIBFKL, ChVeraRGI1} however their practical implementation in phenomenology had been achieved only recently, see e.g.~\cite{ChVeraRGI2, Celiberto:2018muu, Ball:2017otu, Abdolmaleki:2018jln}. For that reason, $k_T$-factorization phenomenology today is still dominated by various heuristic approaches unifying BFKL and DGLAP evolution, such as diferent versions of Catani-Chiafaloni-Fiorani-Marchesini equation~\cite{Jung:2000hk,Hautmann:2013tba,Golec-Biernat:2019scr}, Parton-Branching method~\cite{Martinez:2018jxt}, Kimber-Martin-Ryskin-Watt (KMRW) prescription~\cite{Kimber:2001sc, Watt:2003mx, Watt:2003vf}, Collins-Ellis-Bl\"umlein doubly-logarithmic approach~\cite{Collins:1991ty, Blumlein:1995eu} and many more.

  In the present paper we continue development of the technique of NLO calculations in the the gauge-invariant scheme of High-Energy Factorization, based on Lipatov's gauge-invariant Effective Field Theory(EFT) for Multi-Regge processes in QCD~\cite{Lipatov95, LV}. Following Refs.~\cite{Nefedov:dijet, Karpishkov:BB} we call this scheme -- the Parton Reggeization Approach (PRA). The version of Modified Multi-Regge Kinematics (MMRK) approximation for QCD amplitudes with multiple real emissions was proposed in Ref.~\cite{Karpishkov:BB} to justify the use of KMRW UPDFs together with tree-level amplitudes from High-Energy EFT~\cite{Lipatov95, LV}. Below we will write-down the evolution equation for UPDF, based on MMRK-approximation, which besides leading $\log(1/z)$-terms allows one to resum a subset of subleading logarithmic and $O(z)$ power-suppressed corrections to the UPDF. Then we perform an exploratory NLO calculation for the coefficient function of Deep-Inelastic-Scattering-like subprocess, driven by a gauge-invariant operator ${\rm tr}[G_{\mu\nu}G^{\mu\nu}]$ (where $G_{\mu\nu}$ is a Non-Abelian field-strength tensor), which couples to gluons already in the LO in $\alpha_s$. In CPM, the coefficients functions of this process are known up to $O(\alpha_s^3)$~\cite{Moch:H-DIS} and starting from NNLO they contain doubly-logarithmic terms $\propto \alpha_s^n \log^{2n}(1/z)$ origin of which had been explained in $k_T$-factorization~\cite{Hautmann:2002tu}. Besides inclusive DIS cross-section (or ``structure-function'') we also study the cross-section of production of a leading jet in this process in LO and NLO of PRA. The standard MRK approximation leads to large negative NLO corrections, which for $x_B>10^{-4}$ and $Q\gg 1$ GeV turns NLO cross-section negative, signaling a severe perturbative instability. On the contrary, consistent implementation of our MMRK approximation results in a moderate NLO corrections, which shows, that it solves the major part of the problem of perturbative instability of BFKL formalism at NLO. We trace this results back to the improved treatment of region of initial-state collinear singularity (DGLAP region) in the MMRK approximation. In summary, we have come-up with a practical and manifestly perturbatively-stable recipe of NLO calculations in PRA, which allows one to improve accuracy of the predictions and establish the boundaries of applicability of the approach through the smallness of NLO correction. 

  The paper is organized as follows: in Sec.~\ref{sec:MRK} we formulate the basic formalism of PRA for the particular process we have chosen to study and derive the evolution equation for UPDF in MRK approximation, then in Sec.~\ref{sec:MMRK} we formulate our MMRK approximation, analyze it's performance in comparison to an exact QCD amplitude with one additional emission and write-down UPDF-evolution equation in MMRK approximation and corresponding NLO double-counting subtraction terms; in Sec.~\ref{sec:soft-coll} we describe our phase-space slicing strategy, compute corresponding analytic integrals and double-counting subtraction integral in the soft limit; in Sec.~\ref{sec:virt} we recall the virtual part of NLO-correction under consideration, computed in Ref.~\cite{Nefedov:2019mrg}, and derive corresponding virtual subtraction terms; finally in Sec.~\ref{sec:num} we present and discuss some numerical results and formulate our conclusions. In the \hyperlink{sec:Appendix}{Appendix A} we derive an approximate doubly-logarithmic solution for our UPDF evolution equation, which we use for illustrative numerical calculations throughout this paper and in the \hyperlink{sec:AppendixB}{Appendix B} we take a few iterations of evolution kernel (\ref{eq:C-evol-MRK}) to demonstrate it's properties.    

\section{Basic formalism and UPDF evolution in MRK approximation}
\label{sec:MRK}

  To simplify our presentation, we will always refer to a particular example of hard process -- the DIS-like process (momenta of particles are given in parentheses):
\begin{equation}
{\cal O}(q)+p(P)\to  X, \label{eq:basic-proc}
\end{equation}
 initiated by the gauge-invariant local QCD operator 
\begin{equation}
\lambda{\cal O}(x)=-\frac{\lambda}{2}{\rm tr}\left[G_{\mu\nu}(x)G^{\mu\nu}(x) \right], \label{eq:O(x)}
\end{equation}
  where $\lambda$ is a coupling to an external source and $G_{\mu\nu}=-i\left[D_\mu,D_\nu\right]/g_s$ is a field-strength tensor of QCD with covariant derivative expressed as $D_\mu=\partial_\mu+ig_s A_\mu$ and Hermitian gluon field $A_\mu=A_\mu^a T^a$, where $T^a$ are generators of $SU(N_c)$. In the present paper we will concentrate on the case of pure gluodynamics, i.e. the theory with $n_F=0$. Within Standard Model, the operator (\ref{eq:O(x)}) can be understood as an effective coupling of gluons to a Higgs boson through a loop of very heavy quark, and therefore process (\ref{eq:basic-proc}) can be visualized as a Higgs-exchange contribution to the usual electron-proton DIS. Of course phenomenologically such contribution is negligible, but since the operator (\ref{eq:O(x)}) couples to gluons through the two-gluon vertex:
\begin{equation}
G^{(0),\mu_1\mu_2}_{a_1a_2}=i\lambda\delta_{a_1 a_2}\left( (k_1k_2) g^{\mu_1\mu_2} - k_1^{\mu_2} k_2^{\mu_1}  \right), \label{eq:O-vert}
\end{equation}
where gluon momenta $k_{1,2}$ are incoming and $q+k_1+k_2=0$, as well as through three and four-gluon vertices, proportional to corresponding vertices of QCD with $q+k_1+\ldots+k_3=0$ and $q+k_1+\ldots+k_4=0$, the process (\ref{eq:basic-proc}) have proven to be a useful tool for the formal studies in QCD, probing various aspects of evolution of gluon PDF, see e.g. Refs.~\cite{Moch:H-DIS, Daleo:H-DIS}. 

  In the present paper we will consider the dimensionless inclusive ``structure-function'' of the process (\ref{eq:basic-proc}), which depends on usual DIS kinematic variables $Q^2=-q^2$ and $x_B=Q^2/2(qP)$ and in the LO of CPM is simply equal to
\begin{equation}
F^{\rm (LO\ CPM)}_{\cal O}(x_B,Q^2)= \frac{\pi \lambda^2}{4} x_Bf_g(x_B,\mu_F=Q),\label{eq:F-LO-CPM}
\end{equation}
where $f_g$ is a usual collinear PDF. In our numerical calculations below, we will put the factor $\pi \lambda^2/4=1$. For simplicity, throughout this paper we choose to work in a center-of-momentum frame of $P$ and $q$, where the light-cone components\footnote{For any four-momentum $k$ we define Sudakov decomposition as $k^\mu=\left(n_+^\mu k_- + n_-^\mu k_+ \right)/2+k_T^\mu$ with $k_\pm=n_{\pm}k$, $n_\pm^2=0$, $n_+n_-=2$ and $n_\pm k_T=0$, so that $k^2=k_+k_--{\bf k}_T^2$ and we {\it do not} distinguish covariant and contravariant light-cone components: $k_{\pm}=k^{\pm}$.} of these momenta can be expressed as:
\begin{eqnarray}
P_-=\sqrt{\frac{Q^2}{x_B (1-x_B)}},\ P_+={\bf P}_T=0;\
q_-=-x_B P_-,\ q_+=\frac{Q^2}{x_B P_-},\ {\bf q}_T=0, 
 \label{eq:qP-kinem}
\end{eqnarray} 
i.e. at $x_B\ll 1$ momentum $q$ has large positive (forward) rapidity, while proton flies in the negative direction.

The general expression for inclusive DIS structure-function in CPM is well-known:
\begin{equation}
F_{\cal O}(x_B,Q^2)=  \frac{\pi \lambda^2}{4} \int\limits_{x_B}^1 \frac{dz}{z}\ \frac{x_B}{z}f_g\left(\frac{x_B}{z},\mu_F \right) C(z,Q^2,a_s,\mu_F,\mu^2) + O\left(\left( \Lambda_{QCD}^2/Q^2 \right)^\nu \right),\label{eq:CPM-factor}
\end{equation}  
where the coefficient function $C$ is computed perturbatively as a power-series in $a_s=\alpha_s(\mu^2)/(2\pi)$ and the first (leading-twist) term of Eq.~(\ref{eq:CPM-factor}) is valid up to corrections suppressed as $(Q^2)^{-\nu}$ with $\nu>0$. 

  The $k_T$-factorization hypothesis~\cite{Gribov:1984tu,Collins:1991ty, Catani:1994sq} states, that higher-order corrections to the coefficient function, enhanced by $\log(1/z)$, can be further factorized-out:
\begin{equation}
C(z)= \int \frac{d^2{\bf q}_{T1}}{\pi} \int\limits^{x_B/z}_{x_B} \frac{dx_1}{x_1}\ {\cal C}\left( \frac{zx_1}{x_B},{\bf q}_{T1}, a_s, \mu_F, \mu, \mu_Y \right) H\left(\frac{x_B}{x_1},{\bf q}_{T1},Q^2,a_s,\mu,\mu_Y\right), \label{eq:kT-fact}
\end{equation}
where new coefficient function $H$ is free from potentially large $\log(1/x_1)$-corrections, by $\mu_Y$ we have denoted additional scale which arises due to factorization (\ref{eq:kT-fact}) and spurious dependence on $x_B$ is introduced to make sure, that variable $x_1$ has the same kinematical meaning in Eq. (\ref{eq:kT-fact}) and Eq.~(\ref{eq:kT-fact-F}) below. The statement (\ref{eq:kT-fact}) is proven in QCD in leading-power approximation in $z\ll 1$ (i.e. up to $O(z)$-terms) for the series of Leading-Logarithmic (LL, $\propto [a_s\log^k(1/z)]^n$ with $k=1$ or 2 depeding on a processes)~\cite{BFKL1, BFKL2, BFKL3,Catani:1994sq} and Next-to-Leading Logarithmic (NLL, $\propto a_s[a_s\log^k(1/z)]^n$)~\cite{NLO-BFKL, NLOCiafaloni1, NLOCiafaloni2} corrections. Evolution factor ${\cal C}$ is always single-logarithmic w.r.t. $\log(1/z)$ at leading power in $z$, however additional power of $\log(1/z)$ per $a_s$ can be generated by transverse-momentum integration in Eq. (\ref{eq:kT-fact})~\cite{Hautmann:2002tu}.

  The situation at subleading power is significantly more complicated, see e.g. Refs.~\cite{Chirilli:2018kkw, Bruser:2018jnc}, with doubly-logarithmic corrections arising for some quantities in LLA~\cite{Bartels:1996wc, Ermolaev:2017wym, Penin:2019xql}, however this corrections still can be organized into a sum of terms of a form (\ref{eq:kT-fact}) with different coefficient functions $H$ and evolution factors ${\cal C}$. In the present paper, we work under assumption, that there exist a series of subleading-power ($O(z)$) corrections which can be written in a form (\ref{eq:kT-fact}) {\it with the leading-power coefficient function} $H$, so that all corrections are absorbed into ${\cal C}$, and we assume that this series is numerically dominant for most of inclusive quantities. We make such an assumption, instead of systematically going order-by-order in $z$-expansion, because phenomenologically such an expansion can hardly be expected to be quickly-convergent, since higher-order corrections in QCD typically contain functions like $\log(1-z)$ or $1/(1-z)_+$ with rather slowly-convergent expansion around $z=0$.  

  Substituting Eq.~(\ref{eq:kT-fact}) to Eq.~(\ref{eq:CPM-factor}), changing the order of integrals in $x_1$ and $z$ and making the substitution $z\to zx_B/x_1$ one arrives at a standard High-Energy Factorization formula~\cite{Gribov:1984tu,Collins:1991ty, Catani:1994sq, Nefedov:dijet, Karpishkov:BB}:
\begin{equation}
F_{\cal O}(x_B,Q^2)=\int\limits_{x_B}^1\frac{dx_1}{x_1} \int \frac{d^2{\bf q}_{T1}}{\pi} \Phi_g(x_1,{\bf q}_{T1},\mu,\mu_Y)\times \frac{\overline{|{\cal M}_{\rm PRA}|^2}}{2S x_1} (2\pi)^D \delta(q+q_1-p_{\cal M}) d\Pi_{\cal M}, \label{eq:kT-fact-F}
\end{equation}
where $S=P_-q_+$, $D=4-2\epsilon$, the Unintegrated PDF (UPDF) is
\begin{equation}
\Phi_g(x,{\bf q}_T,\mu,\mu_Y)=\int\limits_x^1 \frac{dz}{z}\ \frac{x}{z}f_g\left(\frac{x}{z},\mu_F \right) {\cal C}\left(z,{\bf q}_T,a_s,\mu_F,\mu,\mu_Y\right),\label{eq:UPDF-def}
\end{equation} 
and we have rewritten the coefficient-function $H$ in terms of ``squared matrix element'' (ME) $\overline{|{\cal M}_{\rm PRA}|^2}$, which at leading power in $z$ is computed as:
\begin{equation}
\overline{|{\cal M}_{\rm PRA}|^2}=\frac{1}{(N_c^2-1){\bf q}_{T1}^2} \left(\frac{q_1^-}{2}\right)^2 |{\cal A}_{\rm EFT}|^2, \label{eq:M2-prescr}
\end{equation}
where ${\cal A}_{\rm EFT}$ is an amputated Green's function of Lipatov's EFT for Multi-Regge processes in QCD~\cite{Lipatov95} with one incoming Reggeized gluon $R_-$ carrying four-momentum $q_1^\mu=q_1^- n_+^\mu/2 + q_{T1}^\mu=x_1P_- n_+^\mu/2 + q_{T1}^\mu$ and some partonic final-state with total four-momentum $p_{\cal M}$. The convention (\ref{eq:M2-prescr}) is introduced (see e.g. Ref.~\cite{Kniehl:2006vm}) to ensure, that in the on-shell limit $|{\bf q}_{T1}|\to 0$ the PRA squared ME reproduces the corresponding squared ME of CPM with Reggeized gluon substituted by an on-shell gluon with four-momentum $n_+^\mu q_1^-/2$, averaged over it's color and helicity of this gluon:
\begin{equation}
 \int\limits_0^{2\pi} \frac{d\phi_1}{2\pi} \lim\limits_{{\bf q}_{T1}\to 0} \overline{|{\cal M}_{\rm PRA}|^2} = \overline{|{\cal M}_{\rm CPM}|^2},\label{eq:coll-lim-M2}
\end{equation} 
where $\phi_1$ is an azimuthal angle of ${\bf q}_{T1}$. Also in Eq.~(\ref{eq:kT-fact-F}) we have introduced a flux-factor $2Sx_1$ which is just a matter of convention, and one have to integrate over the final-state of ${\cal M}$ with the usual Lorentz-invariant phase-space volume element $d\Pi_{\cal M}$.

 The LO PRA subprocess for the process (\ref{eq:basic-proc}) is:
\begin{equation}
{\cal O}(q)+R_-(q_1) \to g(q+q_1), \label{eq:LO-PRA}
\end{equation}
with the following squared ME, derived from vertex (\ref{eq:O-vert}) and LO $R_-\to g$-mixing vertex $\Delta_{-\mu}^{ab}(q)=(-iq^2)n^+_\mu\delta_{ab}$ of EFT~\cite{Lipatov95} (see e.g. Eq. (13) in Ref.~\cite{Nefedov:2019mrg}):
\begin{equation}
\overline{|{\cal M}_{\rm LO}|^2}= \left( \frac{\lambda q_+q_1^-}{2} \right)^2. \label{eq:LO-PRA-ME2}
\end{equation}
 
Substituting Eq.~(\ref{eq:LO-PRA-ME2}) into Eq.~(\ref{eq:kT-fact-F}) one obtains the following expression for the structure function in LO of PRA:
\begin{equation}
F^{\rm (LO\ PRA)}_{\cal O}(x_B,Q^2)= \frac{\pi \lambda^2}{4} \int\limits_0^\infty d{{\bf q}_{T1}^2} \Phi_g(x_1,{\bf q}_{T1},\mu,\mu_Y), \label{eq:F-LO}
\end{equation}
where 
\begin{equation}
x_1=x_B\frac{Q^2+{\bf q}_{T1}^2}{Q^2},\label{eq:x1-LO}
\end{equation}
 due to on-shell condition $(q+q_1)^2=0$ for the final-state gluon. 

\begin{figure}
\begin{center} 
\includegraphics[width=0.7\textwidth]{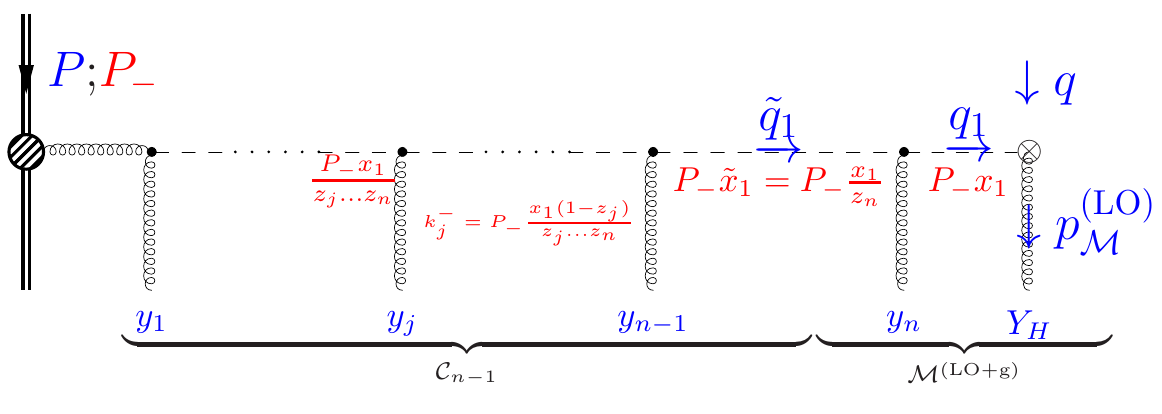}
\end{center}
\caption{UPDF evolution cascade kinematics for the derivation of the MRK evolution equation. Dashed lines denote Reggeized gluons. Vertex with a cross is an insertion of the operator (\ref{eq:O(x)}). \label{fig:ladder}}
\end{figure}

  To set our notation, below we will derive the real-emission term of evolution equation for ${\cal C}$ in the standard MRK approximation, which eventually will coincide with the LO BFKL equation with real emissions ordered in physical rapidity, but rewritten in terms of light-cone momentum fractions $z$ and transverse momenta. To this end let's consider the $k_T$-factorized expression for the contribution to the coefficient function of CPM~(\ref{eq:kT-fact}) with $n-1$ real emissions already factorized into evolution factor ${\cal C}_{n-1}$ and emission of one additional gluon with four-momentum $k_{n}$ in the PRA matrix element (see the Fig.~\ref{fig:ladder}):
\begin{eqnarray}
\frac{d C_{n}(z)}{d\Pi^{\rm (LO)}_{\cal M}}&=& \int\frac{d^2\tilde{\bf q}_{T1}}{\pi} \int\limits_{x_B}^{x_B/z}\frac{d\tilde{x}_1}{\tilde{x}_1} {\cal C}_{n-1}\left(\frac{z\tilde{x}_1}{x_B},\tilde{\bf q}_{T1}\right) \int \frac{dk_{n}^- d^{D-2}{\bf k}_{Tn}}{2k_{n}^-(2\pi)^{D-1}} \int d^D q_1 \delta(\tilde{q}_1-k_{n}-q_1) \nonumber \\
&\times& (2\pi)^D \delta(q_1+q-p^{\rm (LO)}_{\cal M})  \frac{\overline{|{\cal M}_{\rm (LO+g)}|^2}}{2S\tilde{x}_1},\label{eq:Cn-start}
\end{eqnarray}
where $\tilde{q}_1^\mu=(\tilde{x}_1 P_-)n_+^\mu/2 + \tilde{q}^\mu_{T1}$ and we have introduced an intermediate ``$t$-channel'' four-momentum $q_1$. To kinematically factorize-out additional emission one performs the following approximation:
\begin{equation}
\int \frac{dq_1^+ dq_1^-}{2}\ 2\delta(k_{n}^+ + q_1^+)\delta(\tilde{q}_1^--k_{n}^--q_1^-) \times 2\delta({\xcancel{q_1^+}} + q_+ - (p_{\cal M}^{\rm LO})^+) \delta(q_1^- +q_- - (p_{\cal M}^{\rm (LO)})^-),\label{eq:MMRK}
\end{equation}
i.e. neglects the ``small'' light-cone component in the hard process, thus the light-cone components of $q_1$ are: $q_1^+ = -k_{n}^+$, $q_1^-=q_1^- - k_{n}^-$, $\tilde{\bf q}_{T1}={\bf q}_{T1}+{\bf k}_{Tn}$ and one introduces new variables 
\begin{equation}
z_n=q^-_1/\tilde{q}^-_1, \label{eq:zn-definition}
\end{equation}
 and $x_1=q_1^-/P_1^-$, in terms of which, the longitudinal measure of integration becomes:
\[
\frac{d\tilde{x}_1 dk_{n}^-}{\tilde{x}_1 k_{n}^-} = \frac{dx_1}{x_1} \frac{dz_n}{1-z_n},
\]
since $\tilde{x}_1=x_1/z_n$ and $k_n^-=P_- x_1 (1-z_n)/z_n$, see Fig.~\ref{fig:ladder}.

 With approximation (\ref{eq:MMRK}) the $t$-channel momentum transfer in the hard process is equal to:
\begin{equation}
t=q_1^2=-{\bf q}_{T1}^2\left( 1 +  \frac{z_n{\bf k}_{Tn}^2}{(1-z_n){\bf q}_{T1}^2} \right). \label{eq:t_Q-0}
\end{equation}
The approximation (\ref{eq:t_Q-0}) becomes accurate in the limit $Q^2\to 0$. In the Regge limit $z_n\ll 1$ one can put $t\simeq -{\bf q}_{T1}^2$, but the latter approximation quickly degrades with an increase of $z_n$. 
  In the kinematic constraint approach~\cite{Kwiecinski:1996td, Deak:2019wms} to approximately take into account large collinearly-enhanced corrections to BFKL-evolution, one cuts-off the region of real-emission phase-space, where $t\simeq -{\bf q}_{T1}^2$-approximation is no longer valid, i.e. one rejects the emissions with:
   \[
   -q_1^+q_1^- > {\bf q}_{T1}^2 \Leftrightarrow z_n{\bf k}_{Tn}^2 > {\bf q}_{T1}^2(1-z_n).
   \]
   Retaining some realistic approximation for $t$-channel momentum transfer, analogous to Eq.~(\ref{eq:t_Q-0}), will provide the smooth cutoff in the same region of phase-space, thus leading to resummation of essentially the same class of higher-order corrections to BFKL-kernel as in the kinematic constraint approach.  
    
    The squared matrix element with one additional gluon emission, in the Regge limit $z_n\ll 1$ factorizes as follows:
 \begin{eqnarray}
  \overline{|{\cal M}_{\rm (LO+g)}|^2} &=& \left[\frac{1}{\tilde{\bf q}_{T1}^2}\left(\frac{\tilde{q}_1^-}{2} \right)^2\right] \frac{L^2}{(2t)^2} \left[ \frac{1}{{\bf q}_{T1}^2}\left(\frac{q_1^-}{2} \right)^2 \right]^{-1} \times \overline{|{\cal M}_{\rm LO}|^2} \nonumber \\ 
&=& \frac{4C_A g_s^2}{z_n^2 {\bf k}_{Tn+1}^2} \left\{ \frac{{\bf q}_{T1}^2}{t} \right\}^2 \times \overline{|{\cal M}_{\rm LO}|^2}, \label{eq:Mn+1}
\end{eqnarray}
where factors in square brackets are introduced due to normalization prescription~(\ref{eq:M2-prescr}) and the factor $1/(2t)^2$ is the squared tree-level propagator of Reggeized gluon. In the standard MRK-approximation, the factor in curly brackets is put to one as it was discussed above. The factor $L^2=16 C_A g_s^2 (\tilde{\bf q}_{T1}^2{\bf q}_{T1}^2)/{\bf k}_{Tn+1}^2$ is the square of $R_-(\tilde{q}_1)\to R_+(q_1)+g(k_{n+1})$ amputated Green's function in the EFT~\cite{Lipatov95}, i.e. the square of Lipatov's vertex~\cite{BFKL1}:
\[
\Delta^{abc}_{+\mu-}(\tilde{q}_1,q_1)= g_s f^{abc} \left[ -({n}_+{n}_-) \left( (\tilde{q}_1+q_1)_\mu + \tilde{q}_1^2 \frac{{n}^+_\mu}{{q}_1^+} + q_1^2\frac{{n}^-_\mu}{\tilde{q}_1^-} \right)+2\left( \tilde{q}_1^-{n}^+_\mu + {q}_1^+{n}^-_\mu \right) \right].\label{eq:LV}
\] 
Collecting all pieces together, one can rewrite Eq.~(\ref{eq:Cn-start}) as:
\[
\frac{d C_{n}(z)}{d\Pi^{\rm (LO)}_{\cal M}}= \int\frac{d^2{\bf q}_{T1}}{\pi} \int\limits_{x_B}^{x_B/z}\frac{dx_1}{x_1} {\cal C}_n^{\rm (R)}\left(\frac{zx_1}{x_B},{\bf q}_{T1} \right)\times (2\pi)^D \delta(q_1+q-p^{\rm (LO)}_{\cal M})  \frac{\overline{|{\cal M}_{\rm LO}|^2}}{2Sx_1} , 
\]
where 
\begin{equation}
 {\cal C}_{n}^{\rm (R)}(x,{\bf q}_T)= \frac{\alpha_s C_A}{\pi} \int \frac{dz_n}{z_n(1-z_n)} \int\frac{d^{D-2}{\bf k}_{Tn}}{\pi(2\pi)^{-2\epsilon}} \frac{1}{{\bf k}_{Tn}^2} {\cal C}_{n-1}\left(\frac{x}{z_n},{\bf q}_{T1}+{\bf k}_{Tn}\right), \label{eq:subterm_z}
\end{equation}
is the real-emission term of MRK evolution equation, which we have been looking for.

 The limits of $z_n$-integration have to be defined in Eq.~(\ref{eq:subterm_z}). The lower limit is is $x$ due to conservation of $(-)$-component of momentum, while integrating up to $z_n=1$ will lead to non-regularized rapidity divergence due to denominator $1/(1-z_n)$. Technically, the divergence arises because approximation (\ref{eq:MMRK}) violates conservation of $(+)$-momentum component and hence additional emission is allowed to go arbitrarily forward (in the $q$-direction) in rapidity. Demanding, that rapidity of this emission is cut-off at some value $Y_\mu$ one obtains the condition:
\begin{equation}
y_{n}-Y_\mu=\log\left( \frac{|{\bf k}_{Tn}|}{\mu_Y} \frac{z_n}{1-z_n} \right)<0 \Rightarrow z_n<\Delta(|{\bf k}_{Tn}|,\mu_Y),\ \Delta(k_T,\mu)=\frac{\mu}{\mu+k_T}, \label{eq:rap-ord:muY}
\end{equation} 
where function $\Delta$ is familiar from the definition of KMRW UPDF~\cite{Kimber:2001sc, Watt:2003mx, Watt:2003vf} while the rapidity-scale $\mu_Y$ is defined by the relation $q_1^-=\mu_Y e^{-Y_\mu}$ and evolution factor will depend on rapidity scale from now on. In DIS kinematics the good choice for $Y_\mu$ is the rapidity of a parton emitted in the LO PRA subprocess~(\ref{eq:LO-PRA}):
\begin{equation}
Y_\mu \to Y_H := \frac{1}{2}\log\left( \frac{Q^2 (1-x_B)}{{\bf q}_{T1}^2 x_B} \right) \Leftrightarrow \mu_Y \to \frac{Q^2+{\bf q}_{T1}^2}{|{\bf q}_{T1}|}, \label{eq:YH}
\end{equation}
which removes large-logarithmic terms $\propto Y_H$ form coefficient function $H$ at NLO, as we will show in Sec.~\ref{sec:soft-coll}.

  The rapidity of $j$-th gluon in the evolution cascade is given in terms of light-cone momentum fraction of an adjacent $t$-channel parton $z_j$ by:
\begin{equation}
y_j=\log \left(\frac{|{\bf k}_{Tj}|}{P_- x_1}\right) + \log\left(\frac{z_j}{1-z_j}\right) + \sum\limits_{k=j+1}^n\log z_k. \label{eq:yj}
\end{equation}
Applying the latter result to the rapidity-ordering condition for the next emission --  $y_n>y_{n-1}$ one obtains the following choice of rapidity scale for the evolution-factor ${\cal C}_{n-1}$ in the integrand of Eq.~(\ref{eq:subterm_z}): $\mu_Y^{(R,n-1)}\to |{\bf k}_{Tn}|/(1-z_n)$. Thus we have completely obtained all the details of real-emission part of our MRK evolution kernel.

Also, Eq.~(\ref{eq:yj}) allows one to interpret the $z_n$-integration measure in the Eq.~(\ref{eq:subterm_z}) as integration over rapidity $y_n$:
\[
\frac{dz_n}{z_n(1-z_n)}=dy_n.
\]

 To precisely write-down the $D$-dimensional virtual part of the evolution kernel, one can use the one-loop correction to Reggeized gluon propagator (with Born propagator $1/(2t_1)$ factorized-away) of Refs.~\cite{Hentschinski:2011tz, Chachamis:2012cc, Chachamis:2012gh} or, equivalently, the Eq.~(53) in Ref.~\cite{Nefedov:2019mrg} which we reproduce here for the later reference in Sec.~\ref{sec:virt}:
  \begin{eqnarray}
\Pi^{(1)}({\bf q}_{T1}^2,\log r)&=&\frac{\bar{\alpha}_s}{4\pi}\left[ -2C_A (\log r+1) \left(\frac{1}{\epsilon} + \log\frac{\mu^2}{{\bf q}_{T1}^2} \right) \right. \nonumber \\
 &+& \left. \beta_0\left( \frac{1}{\epsilon} + \frac{5}{3} + \log\frac{\mu^2}{{\bf q}_{T1}^2} \right) - \frac{8}{3}C_A + O(\epsilon) \right],\label{eq:R-prop-1}
\end{eqnarray}  
  where $\beta_0=11C_A/3-2n_F/3$, $\bar{\alpha}_s=\mu^{-2\epsilon}g_s^2 (4\pi)^{-1+\epsilon}r_\Gamma$ with $r_\Gamma=\Gamma(1+\epsilon)\Gamma^2(1-\epsilon)/\Gamma(1-2\epsilon)$ and $r\ll 1$ is the parameter of regularization for rapidity divergences in loop integrals, which has been used in Ref.~\cite{Nefedov:2019mrg}. The logarithm $\log(1/r)$ can be identified with the rapidity difference between two adjacent real emissions (see the discussion in the end of Sec.~1 of Ref.~\cite{Nefedov:2019mrg}), and therefore, the virtual part of the evolution equation will be proportional to the coefficient in front of this logarithm -- the one-loop Regge-trajectory of a gluon: 
\begin{equation}
\omega_g({\bf p}_T^2)=-\frac{\alpha_s C_A}{4\pi}\int\frac{d^{D-2}{\bf k}_{T}}{\pi(2\pi)^{-2\epsilon}} \frac{{\bf p}_T^2}{{\bf k}_T^2 ({\bf p}_T-{\bf k}_T)^2} = \frac{\bar{\alpha}_s C_A}{2\pi} \frac{1}{\epsilon} \left(\frac{\mu^2}{{\bf p}_T^2} \right)^\epsilon.\label{eq:omega-g}
\end{equation}

In rapidity-space, one iteration of the real-emission and virtual parts of the evolution kernel has the form~\cite{BFKL1, BFKL2, BFKL3} (see also review~\cite{RevDelDuca95} or a textbook~\cite{kovchegov_levin_2012}):
\begin{equation}
{\cal C}_n(Y_\mu,{\bf q}_T)= \int\limits^{Y_\mu}_{-\infty} dy_n\ \left\{ \frac{\alpha_s C_A}{\pi} \int\frac{d^{D-2}{\bf k}_T}{\pi (2\pi)^{-2\epsilon}} \frac{1}{{\bf k}_T^2} {\cal C}_{n-1}(y_{n},{\bf q}_T+{\bf k}_T) +  2\omega_g({\bf q}_T){\cal C}_{n-1}(y_{n},{\bf q}_T) \right\},\label{eq:BFKL-Y}
\end{equation}
where the factor of two takes into account one-loop contributions from amplitude and complex-conjugate amplitude.

 The starting iteration of the evolution (the LO partonic ``impact-factor'' of the target, in BFKL terminology) is given in rapidity and $x$-space by:
\begin{equation}
{\cal C}_1(y,{\bf q}_T)=\frac{\alpha_s C_A}{\pi} \frac{1}{{\bf q}_T^2} \theta(y-y_1) \Leftrightarrow {\cal C}_1(x,{\bf q}_T,\mu_Y)=\frac{\alpha_s C_A}{\pi} \frac{1}{{\bf q}_T^2} \theta(\Delta(|{\bf q}_T|,\mu_Y)-x), \label{eq:C1}
\end{equation}
and the evolution factor to all orders in $\alpha_s$ in $x$-space is:
\begin{equation}
{\cal C}(x,{\bf q}_T,\mu_Y)={\cal C}_0(x,{\bf q}_T) + \sum\limits_{n=1}^\infty {\cal C}_n(x,{\bf q}_T,\mu_Y), \label{eq:C-series}
\end{equation}
where  ${\cal C}_0(x,{\bf q}_T)=\pi \delta(x-1)\delta({\bf q}_{T1})$ is the perturbative initial condition. 

 As we already have found in Eq.~(\ref{eq:subterm_z}) for the case of real-emission contribution, the Eq.~(\ref{eq:BFKL-Y}) can be equivalently rewritten in terms of light-cone fraction $z$:
  \begin{eqnarray}
{\cal C}_n(x,{\bf q}_T,\mu_Y)&=& \int\limits_x^1 \frac{dz}{z(1-z)} \nonumber \\
 &\times & \left\{ \frac{\alpha_s C_A}{\pi} \int\frac{d^{D-2}{\bf k}_{T}}{\pi(2\pi)^{-2\epsilon}} \frac{1}{{\bf k}_{T}^2} {\cal C}_{n-1}\left(\frac{x}{z},{\bf q}_{T}+{\bf k}_{T},\frac{|{\bf k}_T|}{1-z}\right) \theta\left(\Delta(|{\bf k}_T|,\mu_Y)-z\right)  \right.  \nonumber \\ 
&+& \left. 2\omega_g({\bf q}_T^2){\cal C}_{n-1}\left(\frac{x}{z},{\bf q}_T,\mu_Y\frac{x(1-z)}{z(z-x)}\right)  \theta\left(\Delta(|{\bf q}_T|,\mu_Y)-z\right) \right\}, \label{eq:C-evol-MRK}
\end{eqnarray}
where the particualr rapidity-scale choice in the virtual part: $\mu_Y^{(V,n-1)}=\mu_Yx(1-z)/(z(z-x))$ is uniquely determined by requirements of exponentiation of the virtual part of the evolution and/or cancellation of infra-red divergences between real and virtual parts, as it is shown in the \hyperlink{sec:AppendixB}{Appendix B}. Eq.~(\ref{eq:C-evol-MRK}) is the final form of our MRK evolution equation for the UPDF evolution factor ${\cal C}$.

 It is well-known, that infra-red divergences cancel to all orders in $\alpha_s$ in Eqn.~(\ref{eq:BFKL-Y}) and hence they should also cancel in Eq.~(\ref{eq:C-evol-MRK}) as it is discussed in more detail in the \hyperlink{sec:AppendixB}{Appendix B}. But when one takes the ${\bf q}_{T1}$-convolution of the evolution factor ${\cal C}$ with the coefficient-function $H$ in factorization formula (\ref{eq:kT-fact}), the collinear divergences $\propto (\alpha_s/\epsilon)^n$ are generated to all orders in $\alpha_s$. The latter should be absorbed by usual renormalization of collinear PDF in Eq.~(\ref{eq:UPDF-def}) as it was first systematically done in Ref.~\cite{Catani:1994sq}. It is most convenient to perform this procedure in Fourier-conjugate ${\bf x}_T$-space, because in this space all collinear divergences are contained in the evolution factor ${\cal C}({\bf x}_T)$. We do this in \hyperlink{sec:Appendix}{Appendix A} for the simplified version of Eq.~(\ref{eq:C-evol-MRK}) which strictly neglects all $O(z)$-corrections in the kernel and thus does not depend on the scale $\mu_Y$. The UPDF obtained from doubly-logarithmic solution of this simplified equation is used for illustrative numerical calculations throughout this paper. 
 
\section{Modified MRK approximation: subtraction terms and UPDF evolution}
\label{sec:MMRK}

  To demonstrate the necessity of improving MRK-approximation to reach the stability of NLO corrections in $k_T$-factorization let us consider again, the real-emission NLO correction to the cross-section of the process (\ref{eq:basic-proc}), which is given by PRA subprocess:
  \begin{equation}
  {\cal O}(q) + R_-(q_1)\to g(k_1) + g(k_2). \label{eq:NLO-PRA} 
  \end{equation}
  Let us introduce the convenient variable:
  \begin{equation}
  \hat{z}={k_1^-}/{Q_-},\ Q_-=q^-+q^-_1, \label{eq:zhat-definition}
  \end{equation}
  which together with ${\bf k}_{T1}$, ${\bf k}_{T2}$, $Q^2$ and $x_B$ completely parametrizes {\it exact} $2\to 2$ kinematics of this subprocess. In terms of the latter variable, the contribution of subprocess (\ref{eq:NLO-PRA}) to the SF is given by:
  \begin{equation}
F^{\rm (NLO)}_{\cal O}= \frac{1}{2!} \frac{\pi\lambda^2}{4} \int\frac{d^{D-2}{\bf k}_{T1}d^{D-2}{\bf k}_{T2}}{\pi^2(2\pi)^{-2\epsilon}} \int\limits_0^1 \frac{d\hat{z} }{\hat{z}(1-\hat{z})} w(\hat{z},Q^2,{\bf k}_{T1},{\bf k}_{T2},\alpha_s), \label{eq:CS2-2-w}
\end{equation}
where we have introduced integrand-function -- $w$ and reduced ME -- $f$ which are related with UPDF and squared ME of the subprocess (\ref{eq:NLO-PRA}) as follows:
\begin{eqnarray}
w(\hat{z},Q^2,{\bf k}_{T1},{\bf k}_{T2})&=&\frac{\alpha_s C_A}{\pi}\Phi_g(x_1,{\bf k}_{T1}+{\bf k}_{T2},\mu,\mu_Y) f(\hat{z},Q^2,{\bf k}_{T1},{\bf k}_{T2}), \label{eq:w-exact} \\
f(\hat{z},Q^2,{\bf k}_{T1},{\bf k}_{T2})&=&\frac{\overline{|{\cal M}_{\rm (NLO)}|^2}}{ 4C_A g_s^2\overline{|{\cal M}_{\rm LO}|^2}}, \label{eq:f-def}
\end{eqnarray}
and
\begin{eqnarray}
 Q_-&=&\frac{1}{q_+}\left( \frac{{\bf k}_{T1}^2}{\hat{z}} + \frac{{\bf k}_{T2}^2}{1-\hat{z}} \right),\ k_1^-=Q_-\hat{z},\ k_2^-=Q_-(1-\hat{z}) \label{eq:k1,2m}, \\
 x_1&=&\frac{x_B}{Q^2}\left( Q^2+\frac{{\bf k}_{T1}^2}{\hat{z}} + \frac{{\bf k}_{T2}^2}{1-\hat{z}} \right) . \label{eq:x1-exact}  
\end{eqnarray}

  From Eq.~(\ref{eq:k1,2m}) follows the simple expression for rapidity difference between gluons:
  \begin{equation}
  y_2-y_1=\log\left[ \frac{|{\bf k}_{T2}|}{|{\bf k}_{T1}|} \frac{\hat{z}}{1-\hat{z}} \right], \label{eq:y2-y1} 
  \end{equation}
  which tells us, that for fixed transverse momenta, limit $\hat{z}\to 0$ corresponds to $y_1>y_2$ ($t$-channel Regge limit, $-t/s\ll 1$), while in the limit $\hat{z}\to 1$ one has $y_2>y_1$ ($u$-channel Regge limit, $-u/s\ll 1$). In general, the substitution:
  \begin{equation}
  \hat{z}\leftrightarrow 1-\hat{z},\ {\bf k}_{T1}\leftrightarrow {\bf k}_{T2}, \label{eq:glu_perm}
  \end{equation}
  corresponds to permutation of final-state gluons.
  
  The squared ME of the subprocess~(\ref{eq:NLO-PRA}) can be straightforwardly obtained using Feynman rules of EFT~\cite{Lipatov95} (see e.g. Refs.~\cite{AntonovFRs, Nefedov:2019mrg, Nefedov:dijet, Karpishkov:BB} for the detailed presentation) and is rather long and non-instructive expression, so we refrain from presenting it here. Some relevant limits of it are given below and in Sec.~\ref{sec:soft-coll}.   
  
  As it was shown in Sec.~\ref{sec:MRK} the UPDF-evolution is obtained by factorizing-out an additional gluon emission from subprocess (\ref{eq:NLO-PRA}). Hence, to remove double-counting of Eq.~(\ref{eq:CS2-2-w}) with the evolution, one has to subtract the corresponding approximation for the squared ME from the exact integrand $w$. For standard MRK approximation, the $t$-channel subtraction term is given by Eq.~(\ref{eq:w-exact}) with the following reduced ME:
  \begin{equation}
   f_{{\rm sub.}\ t }^{\rm(MRK)} (\hat{z}, Q^2, {\bf k}_{T1}, {\bf k}_{T2}) = \frac{1}{{\bf k}_{T2}^2}\theta\left(\frac{(1-\hat{z})^2}{\hat{z}^2}-\frac{{\bf k}_{T2}^2}{{\bf k}_{T1}^2} \frac{(Q^2+{\bf k}_{T1}^2)^2}{\mu_Y^2 {\bf k}_{T1}^2}\right), \label{eq:f-sub-t-MRK}
\end{equation}
and UPDF evaluated at $x^{({t},\ {\rm MRK})}_1=x_B\left( Q^2+{\bf k}_{T1}^2/\hat{z} \right)/Q^2$ instead of (\ref{eq:x1-exact}), in accordance with approximation (\ref{eq:MMRK}). The $\theta$-function in Eq.~(\ref{eq:f-sub-t-MRK}) enforces the rapidity-ordering condition (\ref{eq:rap-ord:muY}) and with $\mu_Y=(Q^2+{\bf k}_{T1}^2)/|{\bf k}_{T1}|$ it is equivalent to the condition $y_1>y_2$, see Eq.~(\ref{eq:y2-y1}). The $u$-channel subtraction term is obtained from (\ref{eq:f-sub-t-MRK}) via substitution (\ref{eq:glu_perm}). 

  Clearly, the NLO correction will be smaller if the subtraction term provides a better approximation to an exact squared ME. Since evolution equation is constructed by iterating the same approximation, improvement of the subtraction term will also make the evolution to capture more physics. In fact, as we will see in Sec.~\ref{sec:num}, to obtain meaningful physical results it is crucial to come-up with better approximations to an exact ME, than Eq.~(\ref{eq:f-sub-t-MRK}) can provide. The most important phase-space region, where improvements are necessary, is the DGLAP-region: ${\bf q}_{T1}^2\ll {\bf k}_{T1}^2\simeq {\bf k}_{T2}^2 \ll Q^2$,  integration over which at fixed ${\bf q}_{T1}^2$ generates the contribution enhanced by $\log(Q^2/{\bf q}_{T1}^2)$. The latter large logarithm, when integrated over ${\bf q}_{T1}^2$ with UPDF, leads to sizeable numerical effects.

 \begin{figure}
  \begin{center}
  \includegraphics[width=0.8\textwidth]{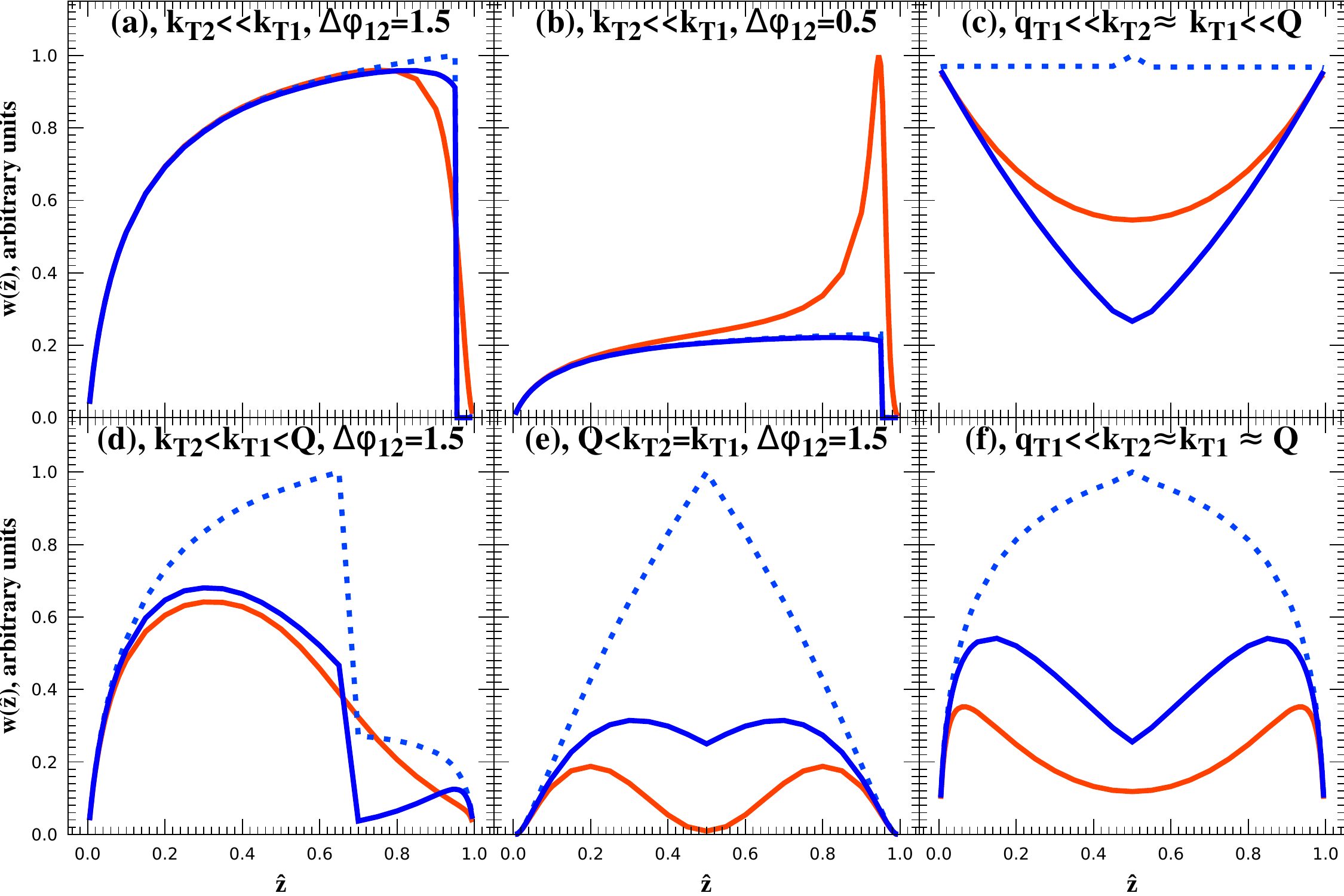}
  \end{center}
  \caption{Plots of integrand function (\ref{eq:w-exact}) as a function of $\hat{z}$. Red solid line -- exact integrand function, dashed line -- standard MRK approximation (\ref{eq:f-sub-t-MRK}), blue solid line -- MMRK approximation (\ref{eq:f-sub-t-MMRK}). Plots (a) -- (f) correspond to different regions of phase-space with different hierarchies of ${\bf k}_{T1,2}$, ${\bf q}_{T1}$ and $Q$. On panels (c) and (f) the function $w$ averaged over azimuthal angle of ${\bf q}_{T1}$ is plotted for the correct on-shell limit (\ref{eq:coll-lim-M2}).  \label{fig:w-plots}}
  \end{figure}
  
  For the squared ME, the DGLAP limit ${\bf q}_{T1}^2\ll {\bf k}_{T1}^2\simeq {\bf k}_{T2}^2 \ll Q^2$ is equivalent to the on-shell limit ${\bf q}_{T1}^2\to 0$ followed by taking ${\bf k}_{T1}^2={\bf k}_{T2}^2\ll Q^2$ asympthotics. Due to Eq.~(\ref{eq:coll-lim-M2}), general (initial-state) collinear factorization theorem for squared MEs in QCD is applicable in this case (see e.g. Eq.~(4.9) of Ref~\cite{Catani:1996vz}) and reduced ME is given by:
   \begin{equation}
   f^{(12)}_{\rm IS-coll.}(\hat{z}, Q^2, {\bf k}_{T2}, {\bf q}_{T1}\to 0) = \frac{\hat{z}p_{gg}(\hat{z})}{2q_1 k_2}, \label{eq:f-IS-coll}
   \end{equation}
   with $2q_1k_2= {\bf k}^2_{T2}/(1-\hat{z})$ and $p_{gg}(z)=z/(1-z)+(1-z)/z+z(1-z)$. Eq.~(\ref{eq:f-IS-coll}) is a non-trivial function of $\hat{z}$, but in Eq.~(\ref{eq:f-sub-t-MRK}) this function is  approximated by a constant. Our goal is to improve this situation leaving the cancellation of infra-red divergences in Eq.~(\ref{eq:C-evol-MRK}) intact. To this end we restore the $({\bf q}_{T1}^2/t)^2$ (``propagator-factor'') in Eq.~(\ref{eq:Mn+1}) with the following approximation for the $t$-channel momentum transfer (compare it with Eq.~(\ref{eq:t_Q-0})):
   \begin{equation}
   t_{\rm MMRK}=-{\bf k}_{T1}^2 - \frac{{\bf k}_{T2}^2 \hat{z}}{1-\hat{z}},
   \end{equation}
   so that reduced ME for the subtraction term in the modified MRK-approximation takes the form:
   \begin{equation}
   f_{{\rm sub.}\ t }^{\rm(MMRK)} (\hat{z}, Q^2, {\bf k}_{T1}, {\bf k}_{T2}) = \frac{1}{{\bf k}_{T2}^2} \left( 1+\frac{\hat{z} {\bf k}_{T2}^2}{(1-\hat{z}) {\bf k}_{T1}^2} \right)^{-2} \theta\left(\frac{(1-\hat{z})^2}{\hat{z}^2}-\frac{{\bf k}_{T2}^2}{{\bf k}_{T1}^2} \frac{(Q^2+{\bf k}_{T1}^2)^2}{\mu_Y^2 {\bf k}_{T1}^2}\right). \label{eq:f-sub-t-MMRK}
   \end{equation}
   
   In the on-shell limit Eq.~(\ref{eq:f-sub-t-MMRK}) reproduces Eq.~(\ref{eq:f-IS-coll}) up to $O(\hat{z}^2)$-terms, so we have partially achieved our goal. In the Fig.~\ref{fig:w-plots} we compare our subtraction terms with an exact integrand function (\ref{eq:w-exact}) numerically. From the plot~\ref{fig:w-plots}(a) one can see, that in the limit $|{\bf k}_{T2}|\ll |{\bf k}_{T1}|$ (or $|{\bf k}_{T1}|\ll |{\bf k}_{T2}|$), both MRK and MMRK subtraction terms approximate an exact integrand very well, except from the region of final-state collinear singularity, which is located at $\Delta\phi_{12}^2+\Delta y_{12}^2\ll 1$ (Fig.~\ref{fig:w-plots}(b)). When all transverse momenta are of the same order, the MRK subtraction term significantly overshoots an exact integrand outside the Regge limits $\hat{z}\to 0$ and $\hat{z}\to 1$ (plots (d) and (e) in the Fig.~\ref{fig:w-plots}), while MMRK expression (\ref{eq:f-sub-t-MMRK}) gives a more reasonable approximation in a whole range of $\hat{z}$. The same behavior is observed in the DGLAP (Fig.~\ref{fig:w-plots}(c)) and on-shell (Fig.~\ref{fig:w-plots}(f)) limits. In general, the MMRK subtraction term is smaller than standard MRK subtraction in whole phase-space, which solves the problem of large negative NLO corrections, typical for BFKL-type calculations, as we will see in Sec.~\ref{sec:num}.   

 Problems which we encounter here are familiar to the practitioners of NLO calculations in High-Energy QCD.  For example, the similar severe over-subtraction problem was observed in the calculation of $p_T$-spectrum of leading forward hadron in proton-nucleus collisions within Color-Glass-Condensate formalism at NLO~\cite{Chirilli:2011km, Chirilli:2012jd, Stasto:2013cha} and was solved in Ref.~\cite{Ducloue:2017mpb} by improvement of the kinematics of the subtraction term. The improvement of MRK approximation for tree-level MEs by propagator factors, as in Eq.(\ref{eq:f-sub-t-MMRK}) is also not new. Such factors where first introduced in the High-Energy Jets (HEJ) approach~\cite{HEJ1, HEJ2}. Also, the new factor, introduced in Eq.~(\ref{eq:f-sub-t-MMRK}) strongly suppresses squared ME in the region which is completely removed in the kinematic constraint approach~\cite{Kwiecinski:1996td, Deak:2019wms}. In Ref.~\cite{Deak:2019wms} it was shown that kinematic constraint approach correctly reproduces certain leading large-logarithmic terms of collinear origin which can be found in the NLO and NNLO expressions for BFKL kernel (the latter is not known in QCD but has been conjectured in ${\cal N}=4$ Supersymmetric Yang-Mils theory~\cite{Caron-Huot:2016tzz}). But e.g. the approach of Ref.~\cite{Altarelli:1999vw} resums the same series of collinear corrections by matching DGLAP and BFKL evolutions. So it seems, that kinematic constraint, MMRK-HEJ and direct resummation approaches are solving the same physical problems of BFKL evolution in a compatible way, but further investigations are needed to confirm this hypothesis.

  Adding the same propagator-factor to real-emission term of evolution equation (\ref{eq:C-evol-MRK}) one ends-up with:
\begin{eqnarray}
{\cal C}_n(x,{\bf q}_T,\mu_Y,\mu_S)&=& \int\limits_x^1 \frac{dz}{z(1-z)} \left\{ \frac{\alpha_s C_A}{\pi} \int\frac{d^{D-2}{\bf k}_{T}}{\pi (2\pi)^{-2\epsilon}} \frac{1}{{\bf k}_{T}^2}  \left( 1+\frac{z {\bf k}_{T}^2}{(1-z) \mu_S^2} \right)^{-2} \right.\nonumber \\ 
&\hspace*{-1.7cm}\times&\hspace*{-1cm}  {\cal C}_{n-1}\left(\frac{x}{z},{\bf q}_{T}+{\bf k}_{T},\frac{|{\bf k}_T|}{1-z},|{\bf q}_T+{\bf k}_T|\right) \theta\left(\Delta(|{\bf k}_T|,\mu_Y)-z\right) \nonumber \\ 
&\hspace*{-1.7cm}+&\hspace*{-1cm}\left.2\omega_g({\bf q}_T^2) {\cal C}_{n-1}\left(\frac{x}{z},{\bf q}_T,\mu_Y\frac{x(1-z)}{z(z-x)},|{\bf q}_T|\right)\theta\left(\Delta(|{\bf q}_T|,\mu_Y)-z\right) \right\}. \label{eq:C-evol-MMRK}
\end{eqnarray}
The MMRK evolution factor depends on an additional scale $\mu_S$, which characterizes the ``hardness'' of the next splitting in the evolution cascade and is an analog of factorization scale in CPM. Technically the scale $\mu_S$ is needed to express kinematical difference between variables $\hat{z}$ (\ref{eq:zhat-definition}) and $z_n$ (\ref{eq:zn-definition}) and for DIS kinematics the optimal choice is $\mu_S^2=Q^2+{\bf q}_{T1}^2$, while the choice $(\mu^{(n-1)}_S)^2=({\bf q}_T+{\bf k}_T)^2$ in the integrand of Eq.~(\ref{eq:C-evol-MMRK}) is due to Eq.~(\ref{eq:t_Q-0}). This equation have to be solved to obtain the UPDF in MMRK-approximation. It might be instructive first to study the ``kinematic constraint'' version of MMRK evolution equation, which is obtained from Eq.~(\ref{eq:C-evol-MMRK}) by the replacement:
\[
\left( 1+\frac{z {\bf k}_{T}^2}{(1-z) \mu_S^2} \right)^{-2} \to \theta((1-z)\mu_S^2 - z {\bf k}_T^2).
\]

  In the end of this section let us make a comment concerning the consistency between real and virtual parts of Eq.~(\ref{eq:C-evol-MMRK}). Our MMRK-approximation has altered only it's real-emission part, leaving the cancellation of IR-divergences intact. The situation here is the same as in the HEJ approach~\cite{HEJ1, HEJ2}, where the standard LO gluon Regge trajectory is used together with modified real-emission amplitudes. However the well-known ``bootstrap'' property of BFKL-equation is lost in Eq.~(\ref{eq:C-evol-MMRK}), i.e. the Regge-factor $\exp\left[ 2\Delta y\ \omega_g({\bf q}_T) \right]$ in not a solution of the color-octet version of it. In our opinion it is an interesting open problem if it is possible to construct the MMRK approximation for real and virtual parts of the evolution equation which would be consistent with the bootstrap.    

\section{Phase-space slicing, soft and final-state collinear integrals, double-counting subtraction in the soft region}
\label{sec:soft-coll}

  To make our NLO calculations more methodologically transparent, we decided to use a simple phase-space slicing method, similar to one proposed in Ref.~\cite{Harris-Owens-NLO}. Our matrix element has non-integrable singularities in two non-overlapping phase-space regions: {\it soft region}, which we define by following cuts on dimensionless energies of gluons:
    \begin{equation}
 \frac{k_1^+ + k_1^-}{q_+}< \delta_s\ {\rm or}\ \frac{k_2^+ + k_2^-}{q_+}< \delta_s,  \label{eq:soft-def}  
  \end{equation}
  where phase-space slicing parameter $0<\delta_s\ll 1$. And (final-state) {\it hard-collinear} region, where:
    \begin{equation}
  \Delta\phi_{1,2}^2 + \Delta y_{1,2}^2 < \delta_c, \label{eq:coll-def}
  \end{equation}
  with $0<\delta_c\ll\delta_s$ and gluons 1 and 2 {\it not} satisfying condition (\ref{eq:soft-def}).
  
  In terms of variables ${\bf k}_{T1,2}$ and $\hat{z}$, the soft condition for the first gluon has the form:
  \begin{equation}
{\bf k}_{T1}^2 < {\bf k}_{T2}^2\hat{z}\left( \delta_s - \hat{z}e^{-2Y_H} \right), \label{eq:soft-cond-k1}
  \end{equation}
  where ${\bf k}_{T2}\simeq {\bf q}_{T1}$, $0<\hat{z}<\hat{z}_{\max}$ with $\hat{z}_{\max}=\delta_s e^{2Y_H}$ (where $Y_H$ is defined in Eq.~(\ref{eq:YH})), and for the second gluon it can be obtained using the substitution (\ref{eq:glu_perm}). 
  
  To facilitate the integration over hard-collinear region, we parametrize ${\bf k}_{T1,2}$ in terms of ${\bf q}_{T1}={\bf k}_{T1}+{\bf k}_{T2}$ and new transverse vector ${\bf \Delta}$ as follows:
  \begin{equation}
  {\bf k}_{T1}=\hat{z} {\bf q}_{T1} + {\bf\Delta},\ {\bf k}_{T2}=(1-\hat{z}) {\bf q}_{T1} - {\bf\Delta},\label{eq:kT12-coll-par}
  \end{equation}
  which in particular allows one to conveniently express the invariant mass of the pair as:
  \begin{equation}
  s=2k_1k_2=\frac{{\bf \Delta}^2}{\hat{z}(1-\hat{z})}.  \label{eq:shat-Delta}
  \end{equation}
  
  In terms of new variable, collinear condition (\ref{eq:coll-def}) has the form:
    \begin{equation}
{\bf \Delta}^2 < {\bf q}_{T1}^2 {\hat{z}}^2 (1-\hat{z})^2 \delta_c,  \label{eq:coll-cond-1}
  \end{equation}
  and requirement of both gluons to be non-soft translates into limits on $\hat{z}$: 
    \begin{equation}
\min(\hat{z},1-\hat{z})>\hat{z}_{\min}=\frac{\delta_s}{1+e^{-2Y_H}}. \label{eq:coll-cond-2}
  \end{equation}
  
  The soft limit of squared PRA amplitude can be computed using the usual eikonal Feynman rule for the emission of a soft gluon with four-momentum $k$ from the hard gluon leg with momentum $p$: $g_s f_{abc} p_i^\mu/(kp_i)$. To take into account the presence of incoming Reggeized gluon $R_\pm$, an additional contribution, proportional to $ (-g_s) f_{abc} n_\mp^\mu/(k_\mp)$ should be added to eikonal amplitude. Hence, for the case at hands, the soft limit is:
  \[
   {\cal M}^{abc,\mu\nu}_{\rm (NLO,soft)} = {\cal M}^{\nu}_{\rm (LO)}\times g_s f^{abc} \left( -\frac{n_+^\mu}{k_1^+} + \frac{k_2^\mu}{(k_1k_2)} \right),
  \]   
  which leads to the following reduced squared amplitude in the $k_1^0\to 0$ soft limit:
   \begin{equation}
f_{{\rm soft}-k_1}=\frac{{\bf k}_{T2}^2\hat{z}^2}{{\bf k}_{T1}^2 ({\bf k}_{T1}-\hat{z}{\bf k}_{T2})^2}. \label{eq:f-soft-1}
 \end{equation}
 Eq.~(\ref{eq:f-soft-1}) has been cross-checked with an exact $D-$dimensional amplitude of the process (\ref{eq:NLO-PRA}) in the soft limit. 
 
 The hard-collinear limit again can be obtained using the standard collinear factorization theorem for squared MEs~\cite{Catani:1996vz}, but this time for final-state singularity:
  \begin{equation}
  f_{\rm FS-coll.}= \frac{p_{gg}(\hat{z})}{s} ,  \label{eq:f-coll}
   \end{equation}
with $s$ given by Eq.~(\ref{eq:shat-Delta}). Eq.~(\ref{eq:f-coll}) also has been verified against an exact squared PRA squared amplitude in $D$-dimensions.
    
   Substituting Eq.~(\ref{eq:f-coll}) to Eq.~(\ref{eq:CS2-2-w}) with the parametrization (\ref{eq:kT12-coll-par}) and cuts (\ref{eq:coll-cond-1}) and (\ref{eq:coll-cond-2}) one finds, that up to effects suppressed as $O(\delta_c, \delta_s)$ the UPDF can be taken out of ${\bf \Delta}$ and $\hat{z}$ integrals:
\[
F^{\rm (coll.)}_{\cal O}= \frac{\pi\lambda^2}{4} \int\limits_0^\infty d{\bf q}^2_{T1}\  \Phi_g(x_1,{\bf q}_{T1},\mu,\mu_Y) H^{\rm (NLO)}_{\rm coll.} ({\bf q}_{T1}, Y_H),
\]
with $x_1$ computed by Eq.~(\ref{eq:x1-LO}) and the following contribution to the NLO coefficient function:
\begin{eqnarray*}
H^{\rm (NLO)}_{\rm coll.}= \frac{1}{2!}\frac{\alpha_s C_A}{\pi} \int\limits_{\hat{z}_{\min}}^{1-\hat{z}_{\min}} d\hat{z}\ p_{gg}(\hat{z}) \int\frac{d^{D-2}{\bf \Delta}}{\pi(2\pi)^{-2\epsilon} {\bf \Delta}^2}\theta\left( {\bf q}_{T1}^2\hat{z}^2(1-\hat{z})^2 \delta_c -{\bf\Delta}^2 \right),    
\end{eqnarray*}
which can be straightforwardly integrated to give:
 \begin{eqnarray}
H^{\rm (NLO)}_{\rm coll.}&=& \frac{\bar{\alpha}_s C_A}{2\pi} \left(\frac{\mu^2}{{\bf q}_{T1}^2}\right)^\epsilon \delta_c^{-\epsilon} \nonumber \\
&\times& \left[ \frac{1}{\epsilon} \left( \frac{11}{6} + 2\log\hat{z}_{\min} \right) - \left( -\frac{67}{9} + \frac{2\pi^2}{3} + 2\log^2(\hat{z}_{\min}) \right) + O(\epsilon) \right], \label{eq:H-coll-final}
 \end{eqnarray}
 where dimensionless coupling $\bar{\alpha}_s$ has been introduced after Eq. (\ref{eq:R-prop-1}).

  Integrating over the soft region (\ref{eq:soft-def}) we can either add up contributions of $k_1^0\to 0$ and $k_2^0\to 0$ limits and divide the cross-section by $2!$, or just integrate Eq.~(\ref{eq:CS2-2-w}) with reduced ME~(\ref{eq:f-soft-1})  over the region (\ref{eq:soft-cond-k1}) and omit the Bose-symmetry factor $1/(2!)$:
  \begin{eqnarray}
H^{\rm (NLO)}_{\rm soft}&=&\frac{\alpha_s C_A}{\pi} \int\limits_0^{\hat{z}_{\max}} \frac{d\hat{z} }{(1-\hat{z})}  \int\frac{d^{D-2}{\bf k}_{T1}}{\pi (2\pi)^{-2\epsilon}} \frac{{\bf q}^2_{T1} \hat{z}}{{\bf k}_{T1}^2 ({\bf k}_{T1}-\hat{z}{\bf q}_{T1})^2} \nonumber \\ &\times& \theta\left({\bf q}_{T1}^2\hat{z}\left(\delta_s-\hat{z}e^{-2Y_H}\right)-{\bf k}_{T1}^2\right).\label{eq:H-soft-0}
\end{eqnarray}
  This integral can be calculated using the well-known Mellin-space representation of the $\theta$-fucntion:
  \begin{equation}
\theta(x-y)=\lim\limits_{\delta\to 0^+}\int\limits_{-i\infty}^{+i\infty}\frac{d\gamma}{2\pi i} \left(\frac{x}{y}\right)^\gamma \frac{1}{\gamma+\delta},\label{eq:theta-Mellin}
\end{equation}   
together with the formula for two-dimensional Euclidean ``bubble'' integral with general indices:
  \begin{equation}
  J_{ab}^\perp({\bf p}_T) = \int\frac{d^{D-2}{\bf k}_T}{({\bf k}_T^2)^a (({\bf p}_T-{\bf k}_T)^2)^b} = \frac{\pi^{1-\epsilon} ({\bf p}_T^2)^{1-a-b-\epsilon} \Gamma(1-a-\epsilon)\Gamma(1-b-\epsilon)\Gamma(a+b+\epsilon-1) }{\Gamma(a)\Gamma(b)\Gamma(2-a-b-2\epsilon)}.\label{eq:perp-master-int}
  \end{equation}
  Also one can notice, that since $\hat{z}<\hat{z}_{\max}\ll 1$, the factor $1/(1-\hat{z})$ in Eq.~(\ref{eq:H-soft-0}) can be omitted up to terms $O(\delta_s)$. Hence after expansion over $\delta_s$ and $\epsilon$ we get:
   \begin{eqnarray}
   H^{\rm (NLO)}_{\rm soft}&=& \frac{\bar{\alpha}_s C_A}{2\pi}\times 2\left(\frac{\mu^2}{{\bf q}_{T1}^2}\right)^{\epsilon} \delta_s^{-2\epsilon} \left( 1-\frac{\pi^2}{6} \epsilon^2 + O(\epsilon^3) \right) \nonumber \\
&\times& \left[ \frac{\xi^{\epsilon}}{\epsilon^2} + \frac{\log(1+\xi)}{\epsilon} + \log\xi \log(1+\xi) - {\rm Li}_2(-\xi)  + O(\epsilon) \right], \label{eq:H-soft}
   \end{eqnarray}
   where $\xi=e^{-2Y_H}$.
   
   Finally, we have to take into account the double-counting subtraction with the evolution. It doesn't influence the collinear limit, since subtraction terms (\ref{eq:f-sub-t-MRK}) or (\ref{eq:f-sub-t-MMRK}) are not singular in the region (\ref{eq:coll-def}) and hence lead to $O(\delta_c)$-suppressed contributions. However, subtraction terms (\ref{eq:f-sub-t-MRK}) or (\ref{eq:f-sub-t-MMRK}) have non-trivial soft limit. The $\hat{u}$-channel subtraction term, which is obtained from Eq. (\ref{eq:f-sub-t-MRK}) or (\ref{eq:f-sub-t-MMRK}) by the substitution (\ref{eq:glu_perm}), in the region (\ref{eq:soft-cond-k1}) reduces to:
   \begin{equation}
   f_{{\rm sub.}\ \hat{u}}^{({\rm soft-}k_1)}= \frac{1}{{\bf k}_{T1}^2}\theta\left(\hat{z}^2 {\bf q}_{T1}^2 \xi_\mu  - {\bf k}_{T1}^2  \right), \label{eq:f-sub-soft-k1} 
   \end{equation}
   where $\xi_\mu=(\mu_Y^2{\bf q}_{T1}^2)/(Q^2+{\bf q}_{T1}^2)^2=e^{-2(Y_H-Y_\mu)}$, see Eq.~(\ref{eq:YH}). One should integrate this expression over region (\ref{eq:soft-cond-k1}) and subtract the result from Eq.~(\ref{eq:H-soft}). We have checked by explicit calculation, that the ``propagator-factor'' in Eq.~(\ref{eq:f-sub-t-MMRK}) makes no difference, up to $O(\delta_s)$-terms, so the double-counting subtraction in the soft limit turns out to be the same for MRK and MMRK approximations. 
   
   Due to a rapidity-ordering $\theta$-function in Eq.~(\ref{eq:f-sub-soft-k1}), we have to split the integration over $\hat{z}$ at a point $\hat{z}_m=\delta_s/(\xi_\mu+\xi)$, so that the subtraction term for the coefficient function takes the form:
  \begin{equation}
  H^{\rm (NLO)}_{\rm sub.} = \frac{\alpha_s C_A}{\pi} \frac{\Omega_{2-2\epsilon}}{ (2\pi)^{1-2\epsilon}} \left[ \int\limits_0^{\hat{z}_m}\frac{d\hat{z}}{\hat{z}} \int\limits_0^{{\bf q}_{T1}^2\hat{z}^2 \xi_\mu} \frac{d{\bf k}_{T1}^2}{({\bf k}_{T1}^2)^{1+\epsilon}} \right. + \left. \int\limits_{\hat{z}_m}^{\delta_s /\xi}\frac{d\hat{z}}{\hat{z}}  \int\limits_0^{{\bf q}_{T1}^2\hat{z}(\delta_s-\hat{z} \xi)} \frac{d{\bf k}_{T1}^2}{({\bf k}_{T1}^2)^{1+\epsilon}} \right], \label{eq:H-sub-0}  
  \end{equation}
  where $\Omega_{2-2\epsilon}=2\pi^{1-\epsilon}/\Gamma(1-\epsilon)$. Calculating this integral, one obtains:
   \begin{equation}
 H_{\rm sub.}^{\rm (NLO)}=\frac{\bar{\alpha}_s C_A}{2\pi} \left(\frac{\mu^2}{{\bf q}_{T1}^2}\right)^\epsilon \left[ \frac{\xi_\mu^{-\epsilon}\hat{z}_m^{-2\epsilon}}{\epsilon^2} + \delta_s^{-2\epsilon} \xi^{\epsilon}\left( \frac{2\log x_0}{\epsilon} -\frac{\pi^2}{3} - \log^2 x_0 + 2{\rm Li}_2(x_0) + O(\epsilon) \right) \right], \label{eq:H-sub-final}
  \end{equation}
  with $x_0=\xi/(\xi_\mu+\xi)$.
  
  Taking all the results of this section together we get:
  \begin{eqnarray}
&&  H^{\rm (NLO)}_{\rm analyt. real}=H^{\rm (NLO)}_{\rm coll.}+H^{\rm (NLO)}_{\rm soft}-H^{\rm (NLO)}_{\rm sub.} = \frac{\bar{\alpha}_s C_A}{2\pi} \left(\frac{\mu^2}{{\bf q}_{T1}^2}\right)^\epsilon \left[ \frac{1}{\epsilon^2} + \frac{1}{\epsilon}\left( \frac{11}{6}+\log\xi_\mu \right) \right. \nonumber \\
&& + \frac{67}{9} - \frac{2\pi^2}{3} +2\log(1+\xi)\left( \log\xi - \log(1+\xi) \right) + \log\delta_c \left(2\log(1+\xi) -\frac{11}{6} - 2\log\delta_s \right) \nonumber \\
&&  -\frac{1}{2}\log\xi_\mu \left( 4\log\delta_s + \log\xi_\mu - 4\log(\xi_\mu+\xi) \right) \nonumber \\
&& \left. -\log^2(\xi_\mu+\xi) -2{\rm Li}_2(-\xi) - 2{\rm Li}_2\left( \frac{\xi}{\xi_\mu+\xi} \right) +O(\epsilon)\right].\label{eq:H-real-analyt}
  \end{eqnarray}
 
  This expression has several important features. First, logarithms of $\delta_s$ in the coefficient in front of $1/\epsilon$, which are present in Eqns.~(\ref{eq:H-coll-final}), (\ref{eq:H-soft}) and (\ref{eq:H-sub-final}) have canceled, giving IR-divergence a chance to cancel against the loop correction. The term $\log\xi_\mu/\epsilon$ will also do so, as we will show in the next section. Second, subtraction (\ref{eq:H-sub-final}) removed all terms proportional to $\log\xi=-2Y_H$, since this logarithm is resummed in the evolution, and only terms decreasing as $e^{-2Y_H}$ are left. And third, if one takes the choice of rapidity scale (\ref{eq:YH}), corresponding to $\xi_\mu=1$, then all terms $\propto \log\xi_\mu$ go away and one is left with:
    \begin{eqnarray}
&& H_{\rm analyt.\ real}^{{\rm (NLO)},\ Y_\mu=Y_H}= \frac{\bar{\alpha}_s C_A}{2\pi} \left(\frac{\mu^2}{{\bf q}_{T1}^2}\right)^\epsilon \left[ \frac{1}{\epsilon^2} + \frac{11}{6}\frac{1}{\epsilon} +\frac{67}{9} - \frac{2\pi^2}{3} \right. \nonumber \\
&&  - \frac{11}{6}\log \delta_c - 2\log\delta_c\log\delta_s + 2\log\delta_c \log(1+\xi) \nonumber \\
&& \left. + 2\log\xi\log(1+\xi) - 2\log^2(1+\xi) +O(\epsilon) \right] . \label{eq:H-real-analyt-no-ximu}
  \end{eqnarray}

  \begin{figure}
  \begin{center}
  \includegraphics[width=0.35\textwidth]{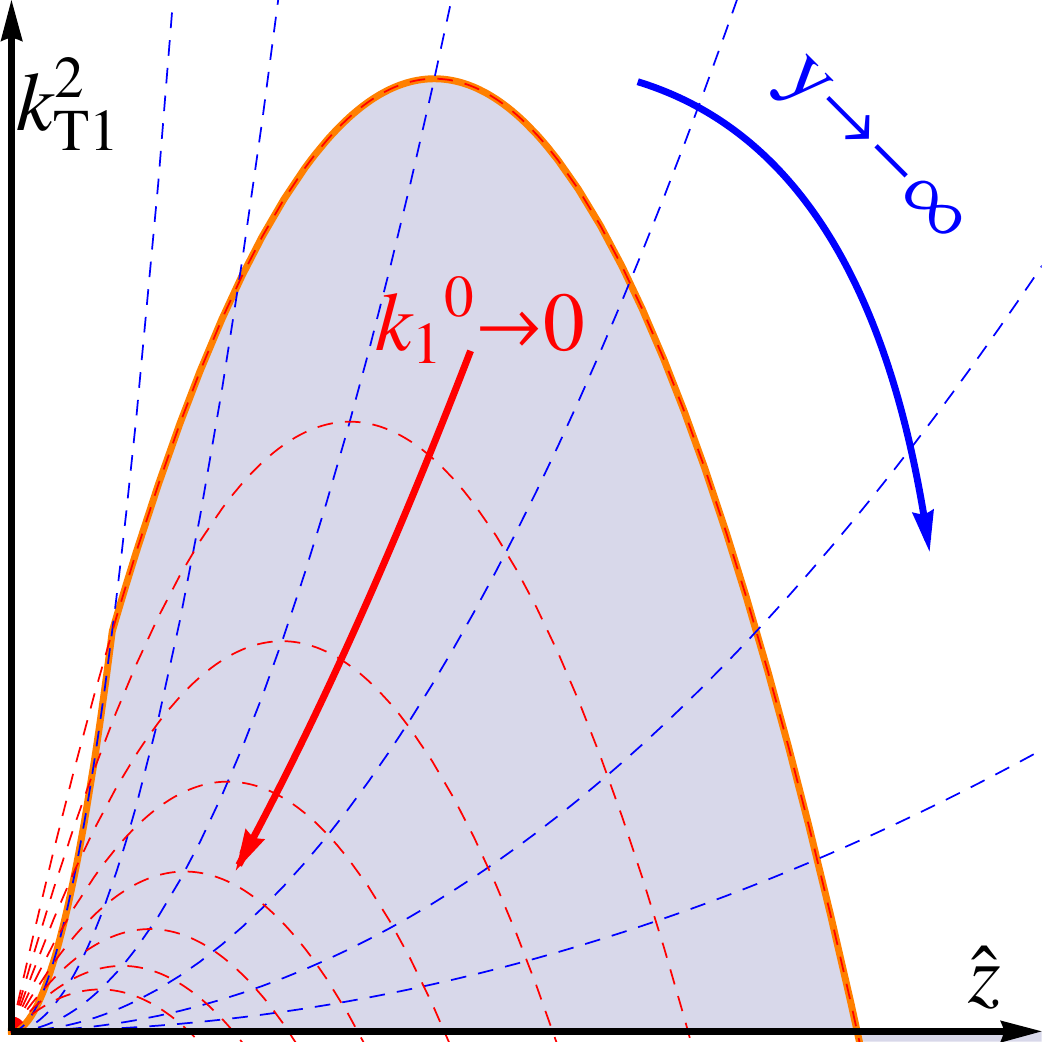}
  \end{center}
  \caption{ The sketch of integration region (shaded area above $\hat{z}$ axis) for soft double-counting subtraction term. Red dashed lines correspond to constant value of $k_1^0/q_+$, while blue dashed lines correspond to constant $y_1$. \label{fig:int-region}}
  \end{figure}
  
  Let us discuss a bit the physical meaning of singularities, arising in the subtraction term (\ref{eq:H-sub-final}). The integration region of Eq.~(\ref{eq:H-sub-0}) is sketched in the Fig.~\ref{fig:int-region} with the lines of constant rapidity of the first gluon and it's constant energy overlaid. 
  
  Going along the line of constant energy in the direction of decreasing rapidity $y_1$ (i.e. in a direction of a proton), one ends-up in the region of small ${\bf k}_{T1}^2$. Hence, the $1/{\bf k}_{T1}^2$-singularity in Eq.~(\ref{eq:f-sub-soft-k1}) is actually a rapidity divergence, corresponding to a fact, that probability of emitting a soft gluon is flat in rapidity. The actual soft divergence is located in a corner $(\hat{z}\to 0, {\bf k}_{T1}^2\to 0)$, where one arrives going along the line of constant rapidity in a direction of decreasing energy. These two divergences overlap in a corner $(\hat{z}\ll \hat{z}_{\max}, {\bf k}_{T1}^2\ll {\bf q}_{T1}^2)$, producing an $1/\epsilon^2$-pole in Eq.~(\ref{eq:H-sub-final}). 
  
  In the soft integral (\ref{eq:H-soft}), the $2/\epsilon^2$-term has two sources. First mechanism is the same overlap of rapidity and soft divergences as in subtraction term, and second -- the overlap of soft and final-state collinear divergences. The $1/\epsilon^2$-pole contribution from the second source, surviving after double-counting subtraction, will cancel against the loop correction.     
  
\section{Virtual correction and subtractions}
\label{sec:virt}

  We have computed the one-loop correction to an amplitude of the process (\ref{eq:LO-PRA}) in Ref.~\cite{Nefedov:2019mrg}. Apart from IR and UV divergences, which we regularize dimensionally, it contains rapidity divergence, physically corresponding to rapidity of a gluon in a loop going far negative. This divergence required additional regularization, which we preform by tilting the Wilson lines in the definition of EFT~\cite{Lipatov95} from the light-cone, as was first proposed in~\cite{Hentschinski:2011tz, Chachamis:2012cc, Chachamis:2012gh}:
  \[
  n_{\pm}^\mu \to \tilde{n}_{\pm}^\mu=n_{\pm}^\mu + r\cdot n_{\mp}^\mu,
  \] 
  where $0<r\ll 1$ is the regularization parameter. As it was already noted in Sec.~\ref{sec:MRK} after Eq. (\ref{eq:R-prop-1}), such regularization roughly corresponds to a smooth cutoff for gluon rapidity at $(-\log r^{-1})/2$ and rapidity divergence manifests itself as $\log r$-term, arising {\it before} one expands loop integrals in $\epsilon$.
  
  The one-loop correction is proportional to the Born vertex~(\ref{eq:O-vert}) (See the last equation in Sec. 4.2 of Ref.~\cite{Nefedov:2019mrg}), and since we are computing $O(\alpha_s)$ correction to the cross-section, we need an interference term:
  \begin{eqnarray}
&&  H^{{\rm (NLO)},\ {\cal O}}_{\rm virt.\ unsubtr.}=2{\rm Re}\left(C\left[ G^{(0)} \right]\right)=\frac{\bar{\alpha}_s}{2\pi} \left\{ -\frac{C_A}{\epsilon^2} + \frac{1}{\epsilon} \left[ \beta_0-C_A(1+L_1) \right] \right. \nonumber \\
&&-C_A\left(\frac{1}{\epsilon} + \log\frac{\mu^2}{t_1} \right)  \log\bar{r}+C_A\left[  2 {\rm Li}_2\left(1-\frac{Q^2}{t_1} \right) +\frac{L_2^2}{2} - L_2 - \frac{1}{2} L_1(L_1+2) + \frac{\pi^2}{6}-\frac{2}{3}  \right] \nonumber\\
&&+\left. \beta_0 \left[ \frac{10}{6}+L_1+L_2 \right] + O(r,\epsilon) \right\}, \label{eq:H-O-unsubtr}
  \end{eqnarray}   
  where $L_1=\log(\mu^2/Q^2)$, $L_2=\log(Q^2/t_1)$, $t_1={\bf q}_{T1}^2$ and $\bar{r}= rQ^2/q_+^2$.
  
  In the Ref.~\cite{Nefedov:2019mrg} we have also shown, that in the full amplitude, which includes one-loop corrections to both scattering vertices and $t$-channel Reggeized gluon propagator (see the right panel of Fig.~5 in Ref.~\cite{Nefedov:2019mrg}), the $\log r$-terms cancel and the (one-Reggeon exchange) EFT result precisely reproduces the (negative-signature part of) the dimensionally-regularized one-loop ($O(g_s^3)$) QCD amplitude of the process 
\begin{equation}
 {\cal O}(q)+g(P)\to g(k_2,Y_H)+g(k_1,y_1), \label{eq:MRK-process}
\end{equation}
 in the Regge limit.

  Here we use this fact to derive the universal subtraction prescription for the virtual rapidity divergence in the one-loop coefficient function (\ref{eq:H-O-unsubtr}), consistent with MRK (\ref{eq:C-evol-MRK}) and MMRK (\ref{eq:C-evol-MMRK}) evolution equations. To this end we start with the interference of one-loop and tree-level corrections to the subprocess (\ref{eq:MRK-process}) which contributes to the CPM coefficient-function of the process (\ref{eq:basic-proc}) in the NNLO($O(\alpha_s^2)$). The EFT predicts leading power Regge ($Y_H-y_1\gg 1$ or $z\to 0$) limit of this interference to be proportional to the squared tree-level matrix element of the subprocess (\ref{eq:MRK-process}) with the one-loop coefficient:
  \begin{eqnarray}
  \frac{2{\rm Re}\left( {\cal M}_{\rm (\ref{eq:MRK-process}),\ 1-loop} {\cal M}^*_{\rm (\ref{eq:MRK-process}),\ tree}\right) }{ \left\vert{\cal M}_{\rm (\ref{eq:MRK-process}),\ tree}\right\vert^2}= H^{{\rm (NLO)},\ {\cal O}}_{\rm virt.\ unsubtr.} ({\bf q}_{T1}^2, Y_H,\log r) \nonumber \\ + H^{{\rm (NLO)},\ g}_{\rm virt.\ unsubtr.}({\bf q}_{T1}^2, y_1,\log r) -2\Pi^{(1)}({\bf q}_{T1}^2,\log r), \label{eq:C-virt-Og}
  \end{eqnarray}   
  where $H^{{\rm (NLO)},\ g}_{\rm virt.\ unsubtr.}$ is the interference of one-loop corrected and tree-level scattering vertices $g(P)\to R_+(q_1)+g(k_2)$ in the EFT, and $\Pi^{(1)}$ is the one-loop correction to the Reggeon propagator (\ref{eq:R-prop-1}). We stress again, that all $\log r$-divergences cancel in Eq. (\ref{eq:C-virt-Og}).

  On the other hand, Eq.~(\ref{eq:C-virt-Og}) already contains large-logarithmic contribution which is reproduced by one iteration of virtual part of evolution equation (\ref{eq:C-evol-MRK}) or (\ref{eq:C-evol-MMRK}) (see e.g. Eq.~(\ref{eq:C2V}) in our \hyperlink{sec:AppendixB}{Appendix B}):
  \[
 (Y^{(2)}_\mu- Y_\mu^{(1)})\times 2\omega_g({\bf q}_{T1}^2),
  \]
  where $Y_{\mu}^{(1,2)}$ are rapidity scales for ${\cal O}R_-g$ and $gR_+g$ scattering vertices, with the optimal choice $Y_\mu^{(1)}\to y_1$, $Y_{\mu}^{(2)}\to Y_H$ and $\omega_g$ is a gluon Regge trajectory~(\ref{eq:omega-g}). Subtracting the latter evolution contribution from Eq.~(\ref{eq:C-virt-Og}) and rearranging the terms, one obtains the following expressions for subtracted one-loop corrections to both scattering vertices:
  \begin{eqnarray}
  &&H^{{\rm (NLO)},\ {\cal O}}_{\rm virt.\ subtr.} ({\bf q}_{T1}^2, Y_H,Y_\mu^{(2)}) = \nonumber \\ && H^{{\rm (NLO)},\ {\cal O}}_{\rm virt.\ unsubtr.} ({\bf q}_{T1}^2, Y_H,\log r) -\Pi^{(1)}({\bf q}_{T1}^2,\log r) -2Y^{(2)}_\mu \omega_g({\bf q}_{T1}^2), \label{eq:HO-virt-subtr-recipie} \\ 
  &&H^{{\rm (NLO)},\ g}_{\rm virt.\ subtr.} ({\bf q}_{T1}^2, y_1,Y_\mu^{(1)}) = \nonumber \\ && H^{{\rm (NLO)},\ g}_{\rm virt.\ unsubtr.} ({\bf q}_{T1}^2, y_1,\log r) -\Pi^{(1)}({\bf q}_{T1}^2,\log r) +2Y^{(1)}_\mu \omega_g({\bf q}_{T1}^2), \label{eq:Hg-virt-subtr-recipie}
  \end{eqnarray}   
  which are also free from $\log r$-divergences. 

By similar reasoning, one can obtain the subtracted one-loop correction to the scattering vertex with any regularization for rapidity divergences, including one proposed in Ref.~\cite{vanHameren:2017hxx}, which opens-up a possibility to automatize the NLO calculations in a variety of small-$x$ physics frameworks. The non rapidity-divergent-part of Eq.~(\ref{eq:R-prop-1}) depends on a chosen rapidity regulator, but the subtracted results (\ref{eq:HO-virt-subtr-recipie}) and (\ref{eq:Hg-virt-subtr-recipie}) should be regularization-independent. 
  
  Using Eq.~(\ref{eq:HO-virt-subtr-recipie}) and subtracting the known (see Ref.~\cite{Moch:H-DIS} and references therein) counter-term for the UV-renormalization of the operator (\ref{eq:O(x)}) in the $\overline{MS}$-scheme: $2\delta Z^{(1)}_{\cal O}= (\bar{\alpha}_s/(2\pi))\beta^{(n_F=0)}_0/\epsilon$  we obtain the following subtracted one-loop coefficient function:
  \begin{eqnarray}
 && H^{{\rm (NLO)},\ {\cal O}}_{\rm virt.\ subtr.} = \frac{\bar{\alpha}_s C_A}{2\pi} \left( \frac{\mu^2}{{\bf q}_{T1}^2} \right)^\epsilon \nonumber \\ &&\hspace*{-10mm} \times \left[ -\frac{1}{\epsilon^2} -\frac{1}{\epsilon}\left( \frac{11}{6}  + \log\xi_\mu \right) + \frac{67}{18} + \frac{\pi^2}{6} + \frac{11}{3}\log\left( \frac{\mu^2}{{\bf q}_{T1}^2}\right) + 2 {\rm Li}_2\left(1-\frac{Q^2}{{\bf q}_{T1}^2} \right)  + O(\epsilon)\right], \label{eq:H-virt-final}
  \end{eqnarray}
 which we have rewritten in terms of $\log \xi_\mu=2(Y_\mu-Y_H)$. The divergence structure in Eq.~(\ref{eq:H-virt-final}) precisely matches that of Eq.~(\ref{eq:H-real-analyt}), so the final answer for analytic part of our NLO correction is:
 \begin{eqnarray}
 && H_{\rm analyt.}^{\rm(NLO)}= H^{{\rm (NLO)},\ {\cal O}}_{\rm virt.\ subtr.} +  H^{{\rm (NLO)}}_{\rm analyt.\ real}= \frac{\bar{\alpha}_s C_A}{2\pi} \left[ \frac{67}{6}-\frac{\pi^2}{2} + \frac{11}{3}\log\left( \frac{\mu^2}{{\bf q}_{T1}^2}\right) + 2 {\rm Li}_2\left(1-\frac{Q^2}{{\bf q}_{T1}^2} \right) \right. \nonumber \\
 && +2\log(1+\xi)\left( \log\xi - \log(1+\xi) \right) + \log\delta_c \left(2\log(1+\xi) -\frac{11}{6} - 2\log\delta_s \right) \nonumber \\
&& -\frac{1}{2}\log\xi_\mu \left( 4\log\delta_s + \log\xi_\mu - 4\log(\xi_\mu+\xi) \right) \nonumber\\
&& \left. -\log^2(\xi_\mu+\xi) -2{\rm Li}_2(-\xi) - 2{\rm Li}_2\left( \frac{\xi}{\xi_\mu+\xi} \right) +O(\epsilon)\right].\label{eq:H-analyt} 
 \end{eqnarray}
 
 The remaining $\mu$-dependence in the coefficient cancels against the running of a coupling $\lambda$ of ${\cal O}(x)$ to an external source. The $\xi_\mu$-dependence should cancel with the UPDF evolution, and with the rapidity-scale choice (\ref{eq:YH}) one obtains:
 \begin{eqnarray}
 && H_{\rm analyt.}^{{\rm(NLO)},\ Y_\mu=Y_H}=  \frac{\bar{\alpha}_s C_A}{2\pi} \left[ \frac{67}{6}-\frac{\pi^2}{2} + \frac{11}{3}\log\left( \frac{\mu^2}{{\bf q}_{T1}^2}\right) + 2 {\rm Li}_2\left(1-\frac{Q^2}{{\bf q}_{T1}^2} \right) \right. \nonumber \\
&&  - \frac{11}{6}\log \delta_c - 2\log\delta_c\log\delta_s + 2\log\delta_c \log(1+\xi) \nonumber \\ &&\left. + 2\log\xi\log(1+\xi) - 2\log^2(1+\xi) +O(\epsilon) \right] .\label{eq:H-analyt-no-ximu} 
 \end{eqnarray} 

For further reference we also write-down the $\delta_{s,c}$-independent part of ${\bf q}_{T1}^2\ll Q^2$-asymptotics of Eq.~(\ref{eq:H-analyt-no-ximu}):
\begin{equation}
H_{{\bf q}_{T1}\to 0}^{{\rm(NLO)}}= \frac{\bar{\alpha}_s C_A}{2\pi} \left[ \frac{67}{6}-\frac{\pi^2}{2} + \frac{11}{3}\log\left( \frac{\mu^2}{{\bf q}_{T1}^2}\right) - \log^2\left( \frac{Q^2}{{\bf q}_{T1}^2} \right) \right], \label{eq:H-asy}
\end{equation} 
which contains very strong negative doubly-logarithmic contribution at small ${\bf q}_{T1}$. We will discuss numerical implications of this in the next section.
  
\section{Numerical results}
\label{sec:num}

  The analytic part of NLO correction, obtained in the previous section, should be added to the numerical integral (\ref{eq:CS2-2-w}) evaluated in $D=4$ space-time dimensions over the region of phase-space where neither condition (\ref{eq:soft-def}) nor condition (\ref{eq:coll-def}) is satisfied and with subtractions (\ref{eq:f-sub-t-MRK}) (MRK) or (\ref{eq:f-sub-t-MMRK}) (MMRK) included at the integrand level. Then, for sufficiently small values of $\delta_c\ll\delta_s\ll 1$, the dependence on this parameters is guaranteed to cancel. In the present section we will show some results of exploratory numerical calculations, performed with UPDF described in the \hyperlink{sec:Appendix}{Appendix A}. Throughout this section we use the scale-choice $\mu_F=\mu=Q$. Our numerical calculations have been performed with the help of parallel version of the well-known \texttt{VEGAS} adaptive Monte-Carlo integration algorithm, implemented in the \texttt{CUBA} library~\cite{CUBA}. The main purpose of this section is to show, that MMRK subtraction term (\ref{eq:f-sub-t-MMRK}) indeed leads to NLO correction with more reasonable physical behavior than MRK subtraction term (\ref{eq:f-sub-t-MRK}).
  
\begin{figure}
  \begin{center}
  \hspace*{-0.5cm}\includegraphics[width=0.9\textwidth]{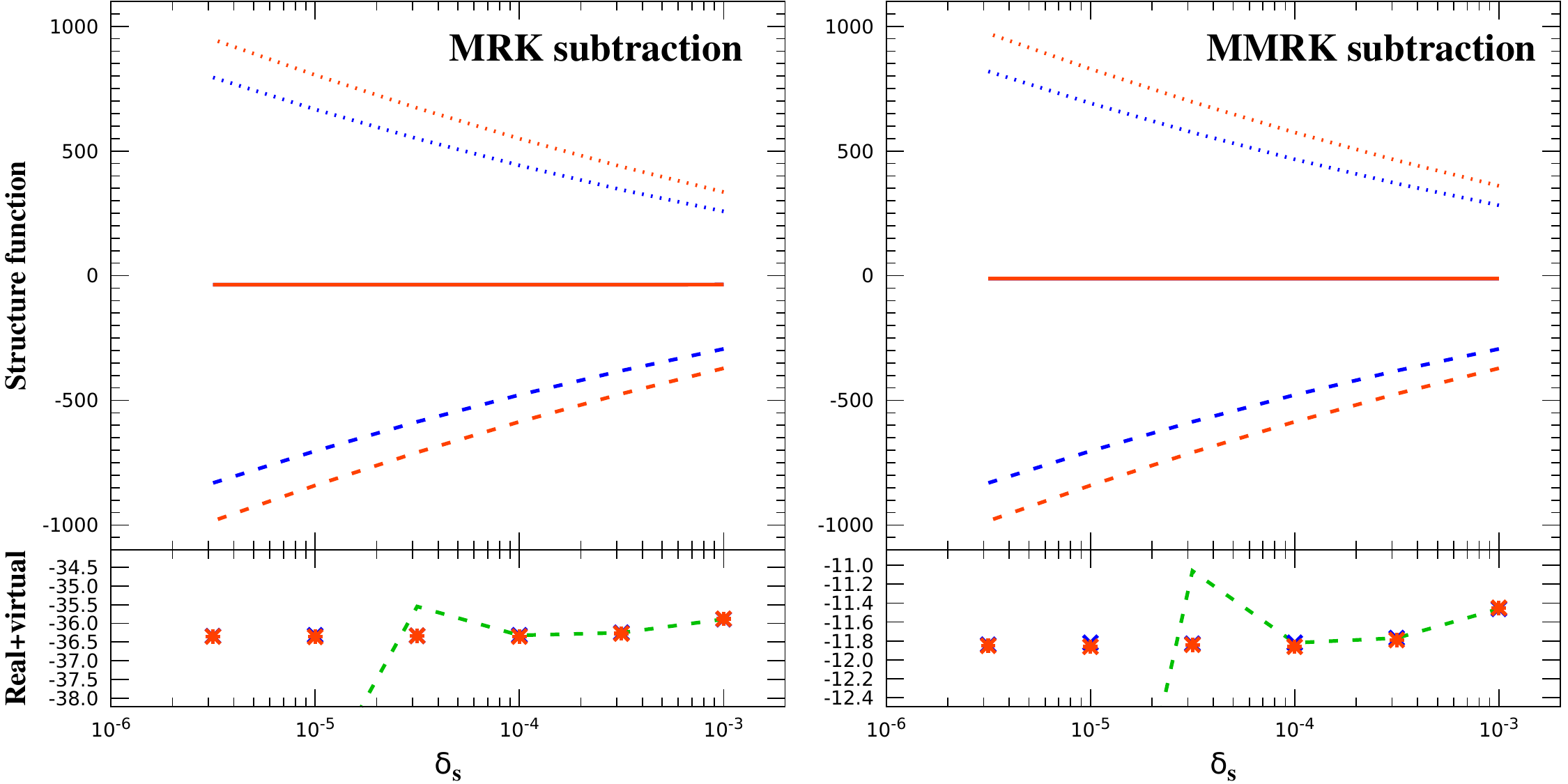}
  \end{center}
  \caption{Test of stability of the NLO PRA correction to the structure function at $x_B=10^{-3}$ and $Q=50$ GeV w.r.t. variation of phase-space slicing parameters $\delta_s$ and $\delta_c$. Dotted lines -- numerical part of NLO correction, dashed lines -- analytic part, obtained with the coefficient function (\ref{eq:H-analyt-no-ximu}), solid line -- their sum. Blue lines correspond to $\delta_c=3\times 10^{-3}\delta_s$, red lines --  $\delta_c=10^{-4}\delta_s$. Left panel -- MRK subtraction (\ref{eq:f-sub-t-MRK}) in the numerical part, right panel -- MMRK subtraction (\ref{eq:f-sub-t-MMRK}). Relative integration accuracy is $10^{-4}$. Quadruple precision arithmetic is used in the evaluation of exact squared ME. The green dashed line in the magnified plots shows the results obtained with double precision.  \label{fig:ds-dc}}
  \end{figure}

  The cancellation of $\delta_s$ and $\delta_c$ dependence is demonstrated in the Fig.~\ref{fig:ds-dc}. One can see, that the result for NLO correction is independent on $\delta_c$ within integration accuracy practically for all points and the plateau in $\delta_s$ is reached rather quickly both for MRK and MMRK subtraction terms. To obtain reliable numerical results at smallest values of $\delta_s$ one have to use quadruple-precision arithmetic in the squared PRA ME subroutine, however such calculations are rather computationally-costly. In the calculations below, we will use fixed values of $\delta_s=2\times 10^{-4}$ and $\delta_c=10^{-3}\delta_s$ with which double-precision calculation is sufficiently stable and accuracy of the NLO cross-section better than $1\%$ is reached. 
  
  \begin{figure}
  \begin{center}
  \includegraphics[width=0.45\textwidth]{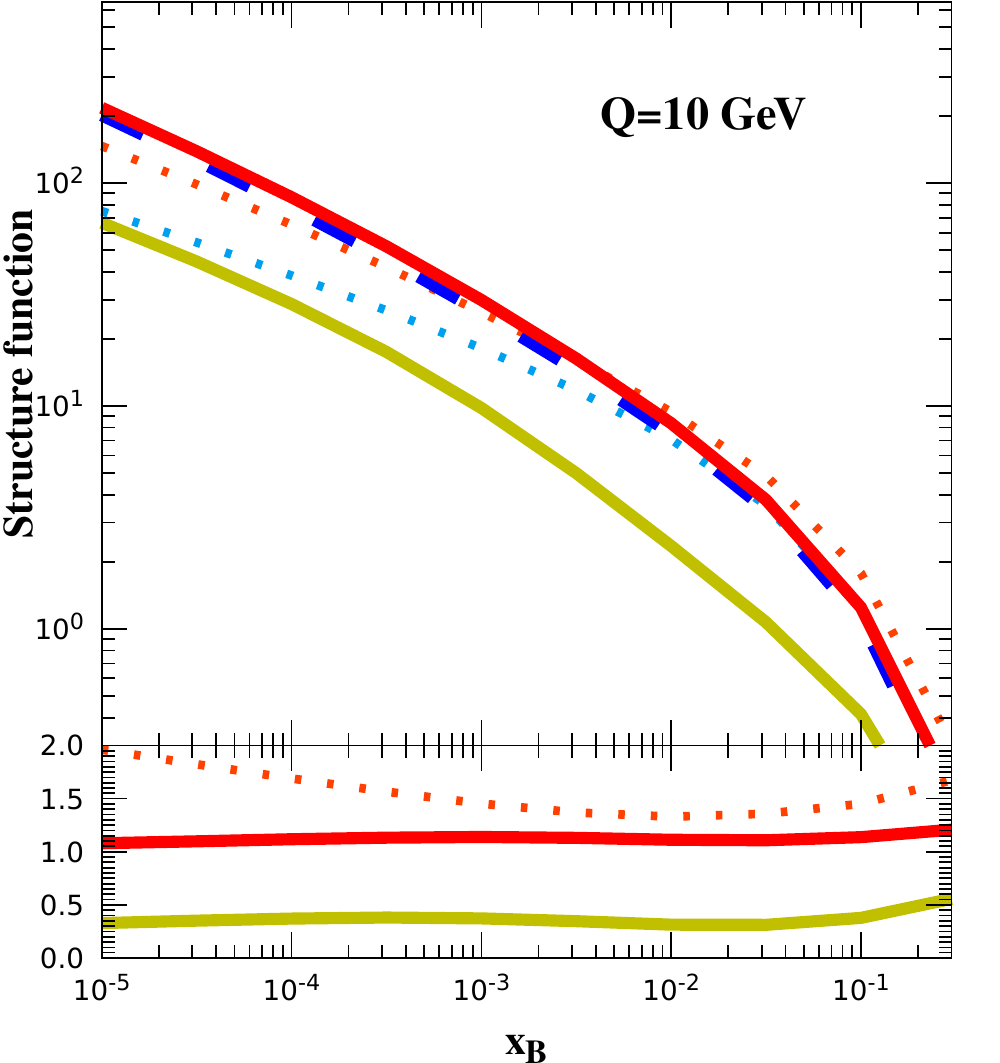} \hspace{0.1cm}
  \includegraphics[width=0.45\textwidth]{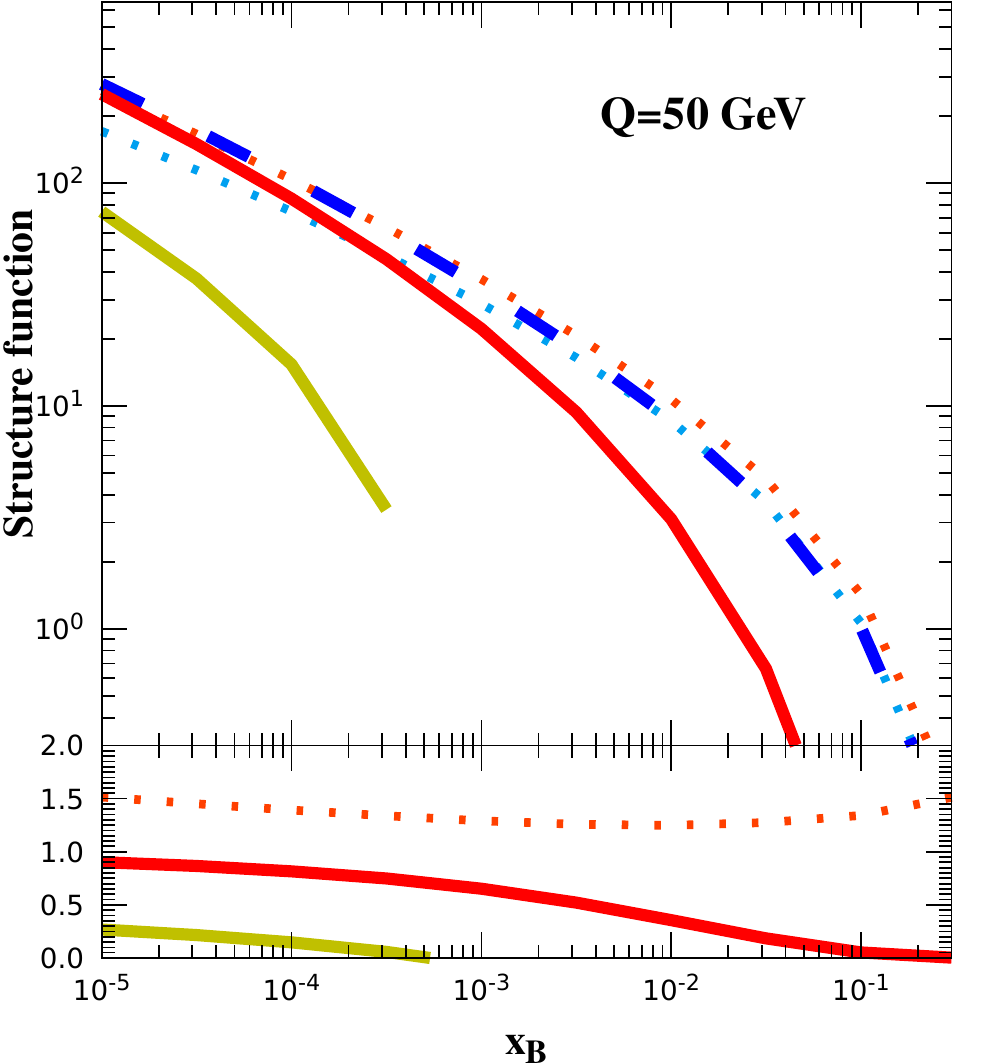}
  \end{center}
  \caption{Inclusive structure function of the process (\ref{eq:basic-proc}) as function of $x_B$ for $Q=10$ GeV (left panel) and $50$ GeV (right panel). Dashed line -- LO PRA (\ref{eq:F-LO}), solid lines: yellow -- NLO PRA with MRK subtraction term, red -- NLO PRA with MMRK subtraction. Dotted lines: blue -- LO (\ref{eq:F-LO-CPM}) and orange -- NLO~\cite{Moch:H-DIS} of CPM. The CPM and PRA NLO/LO ratios (K-factors) are also shown. \label{fig:F-xB}}
  \end{figure}

\begin{figure}
\begin{center}
\includegraphics[width=0.45\textwidth]{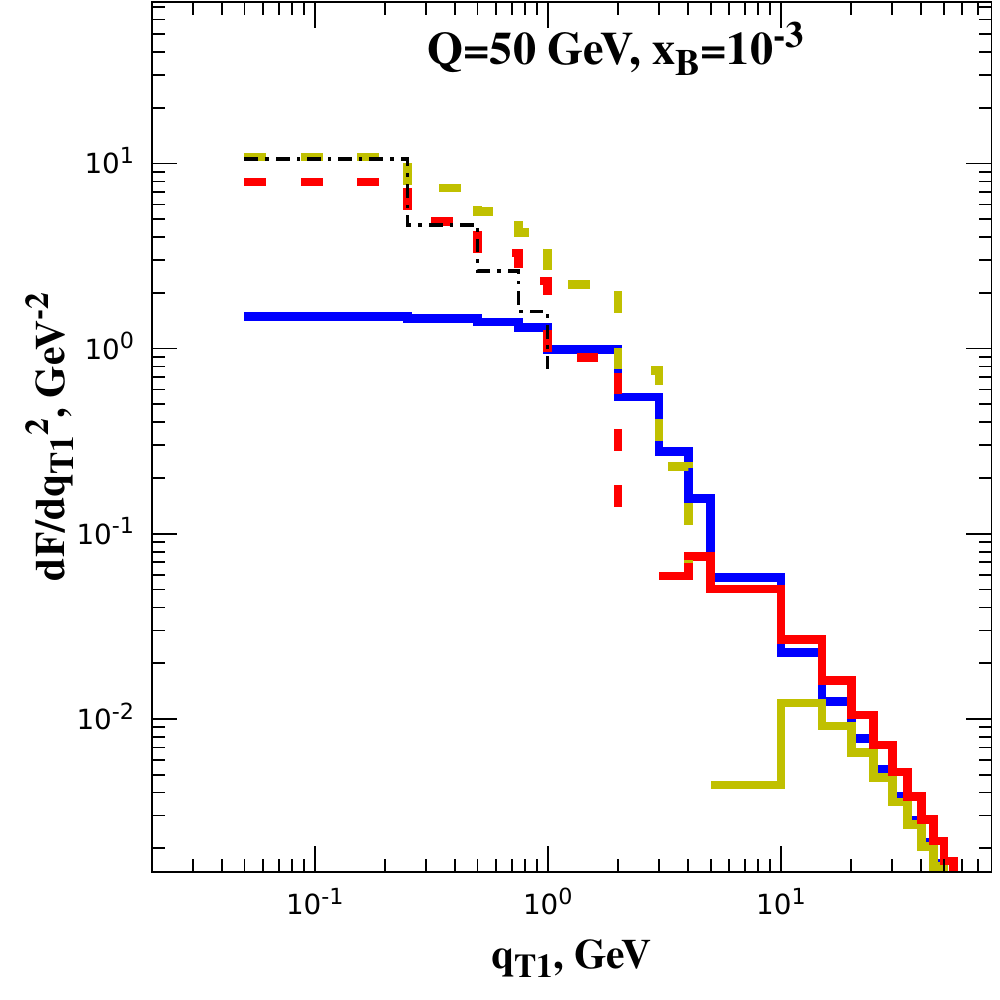}\hspace{0.1cm}
\includegraphics[width=0.45\textwidth]{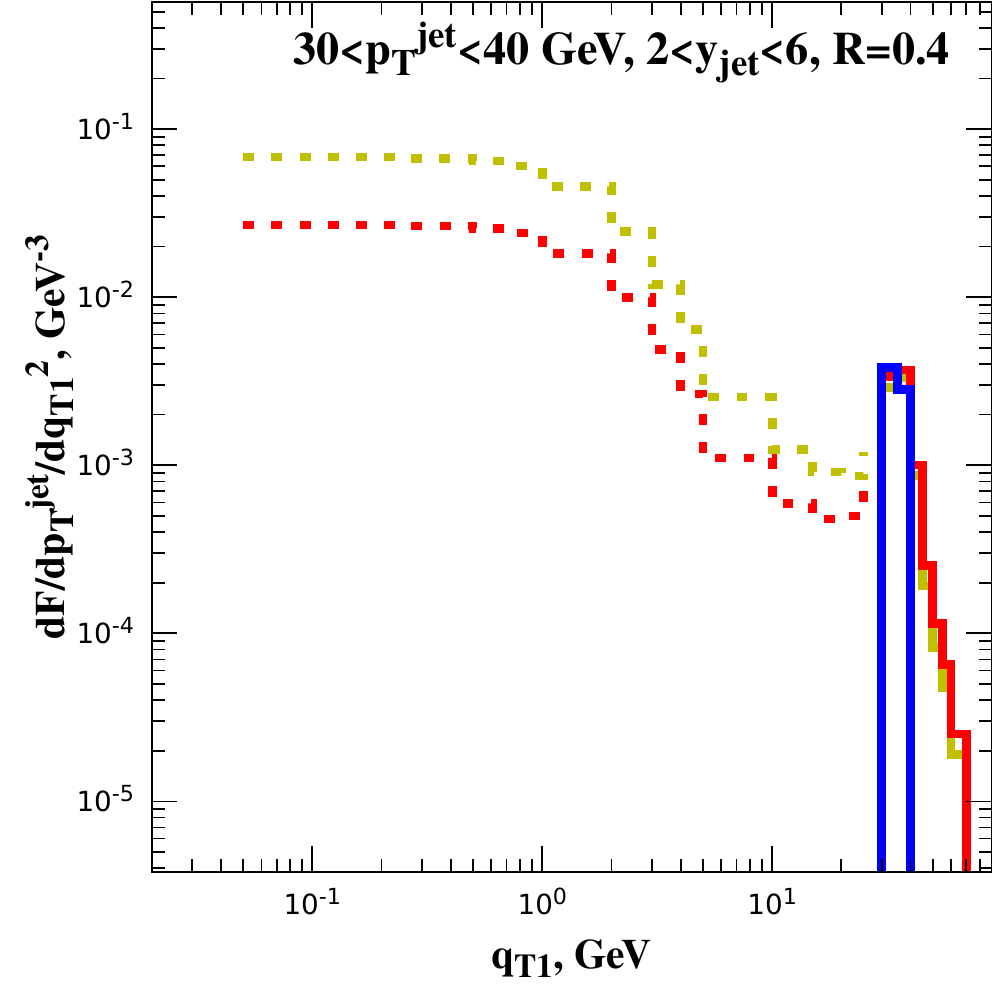}
\end{center}
\caption{ Distribution of NLO correction over $|{\bf q}_{T1}|$ for inclusive SF (left panel) and $30<p_T^{\rm jet}<40$ GeV bin of the jet-$p_T$ spectrum (right panel). Blue solid line -- LO PRA (\ref{eq:F-LO}), yellow lines -- NLO PRA (LO + $O(\alpha_s)$-correction) with MRK subtraction, red lines -- with MMRK subtraction. Dashed histograms correspond to negative values of cross-section. Thin dash-dotted line corresponds to LO PRA curve multiplied by $-(1+H^{\rm (NLO)}_{{\bf q}_{T1}\to 0})$ of Eq.~(\ref{eq:H-asy}). \label{fig:qT12-distr}}
\end{figure}
  
  Next, let us examine the relative size of NLO correction to inclusive structure function with MRK and MMRK subtractions at various $x_B$ and $Q$ (Fig.~\ref{fig:F-xB}). The NLO CPM curve is also shown in the Fig.~\ref{fig:F-xB} and NLO CPM $K$ factor increases towards small $x_B$ and reaches up to a factor of two for $Q=10$ GeV. The NNLO correction in CPM is negative at small $x_B$ and subtracts almost entire NLO correction, while N${}^3$LO correction~\cite{Moch:H-DIS} is positive again and has the same order of magnitude. This kind of behaviour demonstrates lack of stability of CPM calculation at small $x_B$.  The LO PRA curve in Fig.~\ref{fig:F-xB} grows towards small $x_B$ with the same rate as NLO of CPM, calculated with the same collinear PDF, which was used to generate the doubly-logarithmic UPDF. However, this kind of rapid growth is probably an artifact of doubly-logarithmic approximation, which produces UPDF with nonphysically hard high-${\bf q}_T$ tail. With the solution of the full MMRK evolution equation (\ref{eq:C-evol-MMRK}) the small-$x_B$ growth of the structure-function will be significantly slowed-down. However the detailed ${\bf q}_T$-shape of UPDF does not really affect the relative size of NLO correction. We also do not show the scale-variation bands, since doubly-logarithmic UPDF does not depend on scale $\mu_Y$ and corresponding logarithms in the NLO correction will have nothing to cancel against. The more detailed numerical study will be performed once the solution of Eq.~(\ref{eq:C-evol-MMRK}) will become available. 
  
  For both values of $Q=10$ and $50$ GeV, the NLO PRA correction with MRK subtraction is negative and for $Q=50$ GeV it becomes larger than LO PRA term at $x_B>10^{-4}$, demonstrating a severe perturbative instability of the calculation with MRK-subtraction. On the contrary, MMRK-subtracted NLO results look reasonable for $Q=10$ GeV over the whole range of $x_B<0.1$ and for $Q=50$ GeV, the NLO correction becomes larger than 50\% only at $x>10^{-2}$. The $K$-factor of the MMRK-caclculation flattens-out towards small values of $x_B$ as one would expect to see for $\log (1/z)$-resummed calculation, unlike the $K$-factor of CPM calculation. 
  
  It is interesting to investigate the reason, why NLO correction to the inclusive structure function for moderate values of $x_B>10^{-2}$ becomes larger at higher scales. In the left panel of the Fig.~\ref{fig:qT12-distr} we plot the inclusive structure function differential w.r.t. transverse momentum of an incoming Reggeon ${\bf q}_{T1}$ at LO and NLO of PRA. At NLO we find a large negative contribution to the cross-section at small values of $|{\bf q}_{T1}|\ll Q$. This contribution originates from the doubly-logarithmic term $\sim -\log^2(Q^2/{\bf q}_{T1}^2)$ in Eqns.~(\ref{eq:H-analyt-no-ximu}) and (\ref{eq:H-asy}), as it is evident from comparison of the full NLO PRA results with approximate result obtained with an asymptotic coefficient function (\ref{eq:H-asy}) (dash-dotted histogram in the left panel of the Fig.~\ref{fig:qT12-distr}). These doubly-logarithmic effects arise due to exchange of collinear {\it virtual} gluons in the PRA coefficient fucntion and can be factorized from it into a universal Sudakov-type doubly-logarithmic formfactor which will become a part of UPDF. An interesting feature of our formalism is, that this kind of doubly-logarithmic effects is entirely due to virtual exchanges and all doubly-logarithmic effects due to real emissions already had been factorized-out.  We will discuss this problem in more detail elsewhere. 

 In fact, most of the cross-section is accumulated at moderate values of $1\ {\rm GeV}<|{\bf q}_{T1}|<Q$, where doubly-logarithmic effects are much less important than single-logarithmic effects which we discuss in the present paper. With $x_B$-decreasing, contribution of moderate and large $|{\bf q}_{T1}|$ grows (see Fig.~\ref{fig:UPDFs} in the \hyperlink{sec:Appendix}{Appendix A}), which explains, why NLO PRA $K$-factor flattens-out towards small $x_B$ and becomes much less sensitive to the value of $Q^2$.   

  Single-logarithmic effects $\sim\log(Q^2/{\bf q}_{T1}^2)$ which drastically improve the quality of MMRK-approximation in comparison with the MRK calculation, come from the DGLAP region of NLO real-emission phase-space ${\bf q}_{T1}^2\ll {\bf k}_{T1,2}^2 \ll Q^2$. In CPM this is a region of initial-state collinear divergence, which is subtracted from NLO correction and governs PDF evolution. In PRA, there is no initial-state collinear divergence in the coefficient-function calculation. All collinear divergences are subtracted at the level of UPDF (see \hyperlink{sec:Appendix}{Appendix A}). But the DGLAP region still generates large contribution, enhanced by $\log(Q^2/{\bf q}_{T1}^2)$ and proportional to the DGLAP splitting function $p_{gg}(\hat{z})$, see Eq.~(\ref{eq:f-IS-coll}). Double-counting subtraction term partially subtracts this contribution, but for MRK-approximation, the quality of this subtraction is very poor, resulting in a huge negative NLO correction. MMRK subtraction term approximates an exact DGLAP splitting function better, as it was discussed in Sec.~\ref{sec:MMRK}, but the remaining mismatch still generates some ``collinear'' logarithmic term. However, at smaller values of $x_B$ and $Q$ this logarithmic term does not present such a big problem as for $x_B$ closer to one and at higher $Q\gg \Lambda$, where $\Lambda\sim 1$ GeV is a scale of non-perturbative transverse momentum, which is present in UPDF. This factor also contributes to better stability of NLO correction at $Q=10$ GeV vs. $50$ GeV in the Fig.~\ref{fig:F-xB}.   
  
  An interesting feature of PRA is, that many observables related with transverse momentum are available already in the LO. For the process at hands, such an observable is a leading jet $p_T$-spectrum, which at LO is just a structure function (\ref{eq:F-LO}), differential in transverse momentum ${\bf q}_{T1}$ and jet rapidity $Y_H$ (\ref{eq:YH}). Therefore now we have an opportunity to study the relative size of NLO correction in PRA also for the jet-$p_T$ spectrum.
  
  To meaningfully define the jet observable at NLO we need to take into account, that UPDF evolution is not ordered in $p_T$, so it can produce jets with transverse momenta higher than $p_T$ of a jet originating from the hard process. But evolution is ordered in rapidity, so we can avoid the need of fully-exclusive Monte-Carlo simulation by reasonably defining the ``most forward'' high-$p_T$ jet. We do this as follows:
  \begin{enumerate}
  \item If both gluons in the NLO subprocess (\ref{eq:NLO-PRA}) have $\Delta y_{1,2}^2+\Delta\phi_{1,2}^2 < R^2$ with jet-radius parameter $R=0.4$ in our numerical calculations below, their four-momenta are added to form a four-momentum of a jet.
  \item Otherwise the four-momentum of a gluon leading in $p_T$ and lying within rapidity-acceptance is taken as a jet four-momentum
  
  \item If rapidity of the gluon subleading in $p_T$ is $y_{\rm subl.}<y_{\rm jet}$, then it is unconstrained
  
  \item If $y_{\rm subl.}>y_{\rm jet}$, we reqire it's $p_T^{\rm (subl.)}<p_T^{\rm (veto)}=10$ GeV.
  \end{enumerate}
  
  In all other respects, our NLO calculation for jet-$p_T$ spectrum proceeds the same way as for inclusive structure function, with no need to re-calculate the analytic part, just the phase-space slicing parameters have to be taken sufficiently small to avoid interference with jet definition. 

\begin{figure}
\begin{center}
\includegraphics[width=0.45\textwidth]{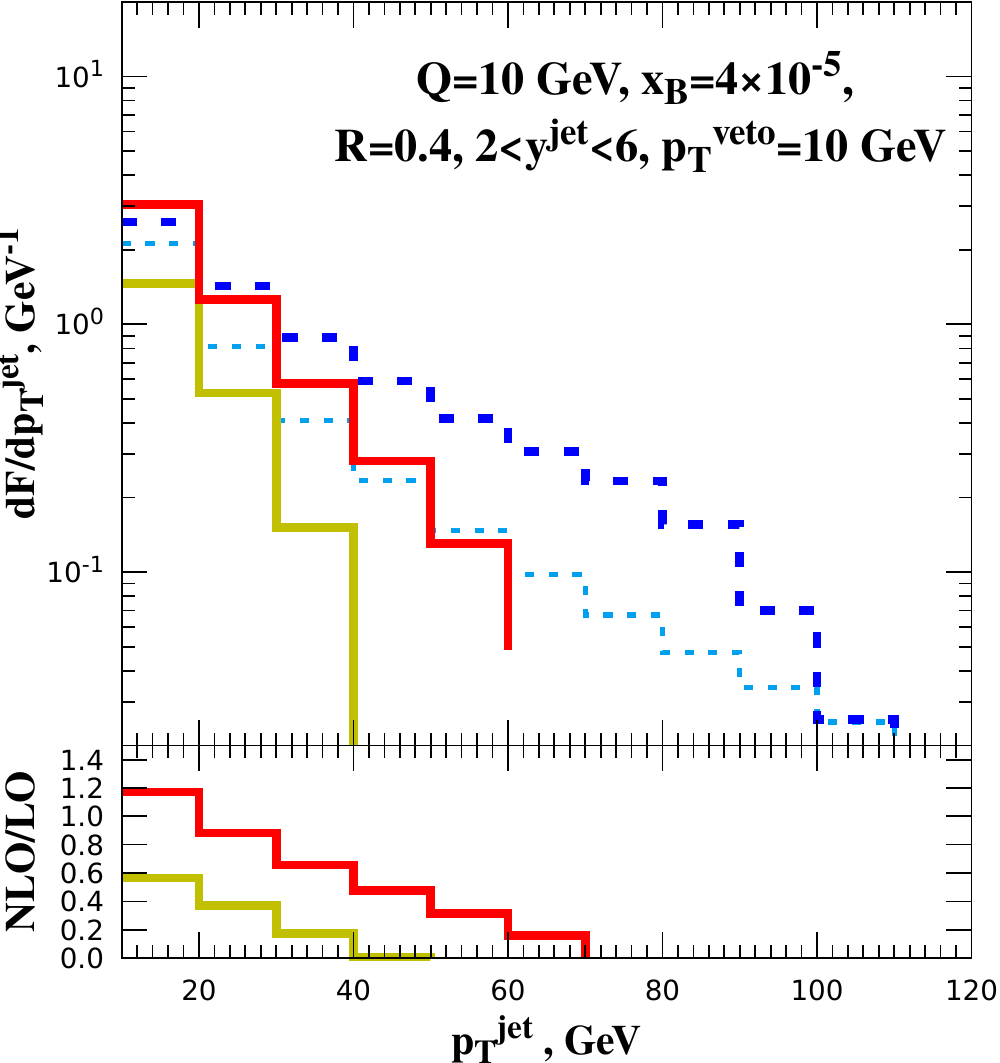}
\includegraphics[width=0.45\textwidth]{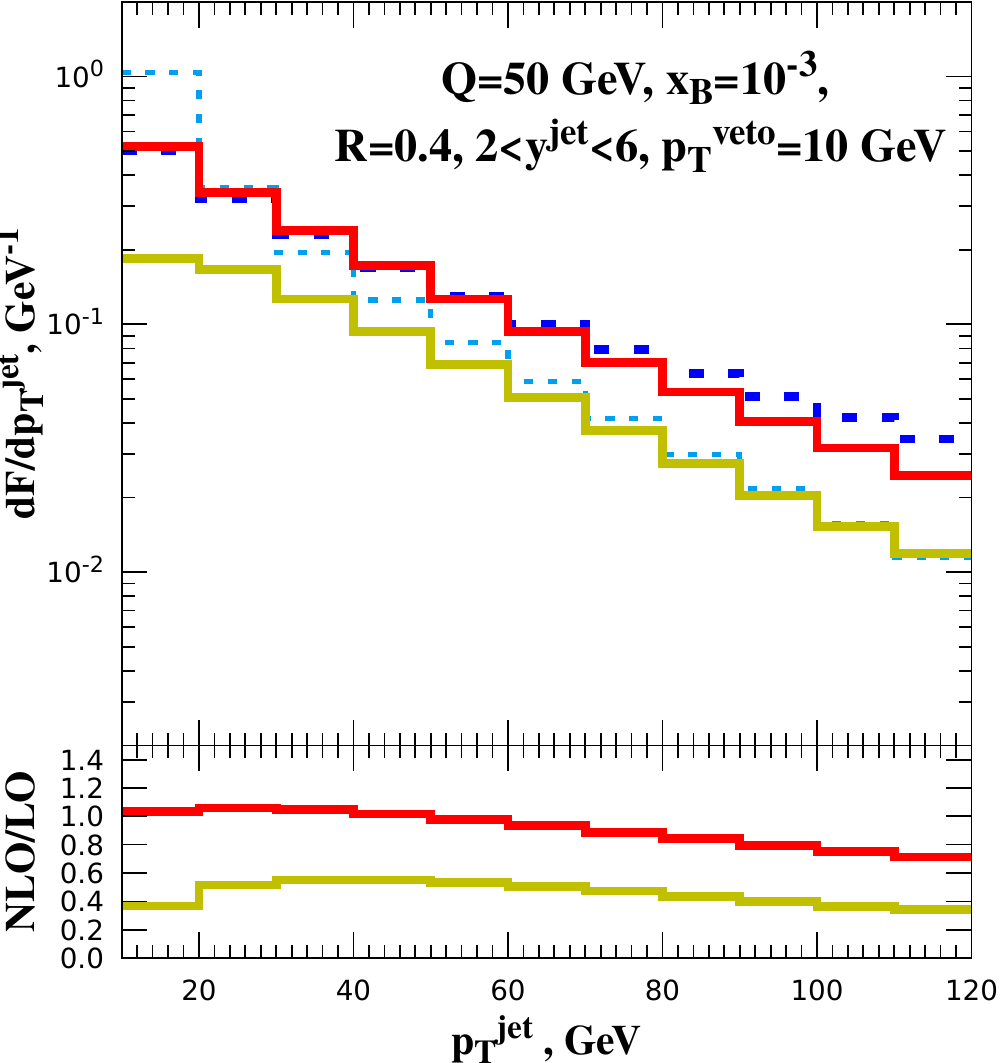}
\end{center}
\caption{ Numerical results for jet-$p_T$ spectrum in the process (\ref{eq:basic-proc}). Dashed line -- LO PRA, dotted line -- LO CPM, solid lines: yellow -- NLO PRA with MRK subtraction, red -- with MMRK subtraction.  \label{fig:jet-pT}}
\end{figure}

  Numerical results for jet-$p_T$ spectrum are shown in the Fig.~\ref{fig:jet-pT} for two different values of $Q=10$ and $50$ GeV and the same value of ``center-of-mass energy'' $S=Q^2/x_B$. For jet $p_T$-spectrum we use the same factorization and renormalization scale-choice as for inclusive SF. Here we find larger NLO corrections at smaller scales, which most likely reflects a steeper decrease of PDF with increasing values of $x$ at smaller scales. The MMRK approximation again leads to smaller NLO correction and at $Q=50$ GeV, the NLO correction to jet-$p_T$ spectrum is negligible in most bins, suggesting that LO PRA is a good approximation for this observable in this kinematic region.   

  Finally, the right panel of the Fig.~\ref{fig:qT12-distr} allows us to examine, how NLO correction to the jet $p_T$-spectrum is distributed w.r.t. transverse momentum of incoming Reggeon ${\bf q}_{T1}^2$. For the jet-$p_T$ spectrum, the loop correction and IR-cancellation effects are present only in two bins in the right panel of Fig.~\ref{fig:qT12-distr} where the LO term is nonzero, and one can see, that NLO result in this bins is very close to LO, so NLO correction in those bins is negligible. It is the behavior of subtraction term, which is responsible for the bulk of NLO correction to the jet-$p_T$ spectrum.

  We emphasise, that doubly-logarithmic effect $\sim \log^2(Q^2/{\bf q}_{T1}^2)$, which was de-stabilizing the NLO correction to the inclusive structure function is absent for the jet $p_T$-spectrum with $p_T^{\rm jet}\sim Q$ so the NLO-correction to this observable is particularly stable and suitable for calculations in $k_T$-factorization and PRA. This conclusion is likely to generalize on a wide class of observables which are sensitive to the transverse momentum of incoming partons and  absent in the LO of CPM. The $\Delta\phi$-spectrum of a dijet or multi-jet system at small values of $\Delta\phi$ (away from back-to-back configuration) is a good example of such an observable~\cite{Nefedov:dijet, Bury:forward-dijet, Kutak:4-jet}.  

  As a conclusion we emphasize, that in the present paper we have formulated the technique of NLO calculations in PRA for gluon-induced processes and the MMRK-approximation for squared matrix element in QCD with emission of additional partons. This approximation should be used consistently as the subtraction term in NLO correction and in the UPDF evolution, leading to improved perturbative stability of the calculation both for inclusive structure function and jet cross-section. 

\section*{Acknowledgments}
 Author is grateful to Krzysztof Kutak, Andreas van Hameren, Bernd Kniehl and Vladimir Saleev for multiple thought-provoking discussions of various aspects of NLO calculations in $k_T$-factorization, to Sven-Olaf Moch for providing the numerical code of Ref.~\cite{Moch:H-DIS} for coefficients functions of the process (\ref{eq:basic-proc}) up to $N^3LO$ in CPM and to the helpful referee, for attracting our attention towards the problem of reconciling collinear corrections with the bootstrap. The work was supported in parts by the Ministry of education and science of Russia through the State assignment to educational and research institutions under project FSSS-2020-0014, the RFBR grant \# 18-32-00060 and by the Foundation for the Advancement of Theoretical Physics and Mathematics BASIS through grant No. 18-1-1-30-1.
  
\section*{Appendix A: UPDF in doubly-logarithmic approximation}  
\hypertarget{sec:Appendix}{}

  To demonstrate how collinear divergences can be subtracted to all orders from the UPDF we will closely follow Ref.~\cite{Catani:1994sq}. Let us simplify Eq.~(\ref{eq:C-evol-MRK}) by omitting all $O(z)$ corrections to the kernel, which in turn leads to disappearance of the $\mu_Y$-scale dependence:
\begin{eqnarray}
\tilde{\cal C}(x,{\bf q}_T)&=&  \delta(x-1)\delta({\bf q}_{T}) \label{eq:BFKL}\\
&+&\frac{{\alpha}_s C_A}{\pi} \int\limits_x^1 \frac{dz}{z} \left\{ \int\frac{d^{D-2}{\bf k}_{T}}{\pi (2\pi)^{-2\epsilon} } \frac{1}{{\bf k}_{T}^2} \tilde{\cal C}\left(\frac{x}{z},{\bf q}_{T}+{\bf k}_{T}\right)\right. + \left. r_\Gamma \frac{(4\pi)^{\epsilon}  ({\bf q}_T^2)^{-\epsilon}}{\epsilon} \tilde{\cal C}\left(\frac{x}{z},{\bf q}_T\right) \right\}\nonumber,
\end{eqnarray}
where $\tilde{\cal C}={\cal C}/\pi$. To facilitate taking the iterations of this kernel and subtraction of collinear divergences, we pass to the Mellin representation for the $x$-dependence of the evolution factor and transverse-position space for it's ${\bf q}_T$-dependence:
\begin{equation}
\tilde{\cal C}(N,{\bf x}_T,\mu)=\int\limits_0^1 dx\ x^{N-1} \int d^{D-2}{\bf q}_T\ e^{i{\bf x}_T{\bf q}_T} \tilde{\cal C}(x,{\bf q}_T,\mu). \label{eq:Mellin-Fourier} 
\end{equation}

  In this representation, Eq.~(\ref{eq:BFKL}) takes the form:
\begin{eqnarray}
&&\tilde{\cal C}(N,{\bf x}_T,\mu) =1 + \frac{\hat{\alpha}_s}{N} \frac{\Gamma(1-\epsilon) (\mu^2)^\epsilon}{(-\epsilon)\pi^{-\epsilon}}\int d^{D-2}{\bf y}_T \nonumber \\ && \times \left[({\bf x}_T^2)^{\epsilon}\delta({\bf x}_T-{\bf y}_T)  - \frac{\epsilon \Gamma(1-\epsilon)}{\pi^{1-\epsilon}} \left(({\bf x}_T-{\bf y}_T)^2 \right)^{-1+2\epsilon}\right]\tilde{\cal C}(N,{\bf y}_T,\mu),
\end{eqnarray}
where $\hat{\alpha}_s={\alpha}_s(\mu) (\mu^2)^{-\epsilon} C_A/\pi$. Taking iterations of this equation reduces to repeated use of Eq.~(\ref{eq:perp-master-int}), so for the $n+1$-th iteration one finds (compare with Eqns.~(3.3) -- (3.5) of Ref.~\cite{Catani:1994sq}):
\begin{eqnarray}
&& \tilde{\cal C}_{n+1}(N,{\bf x}_T,\mu)= \frac{\hat{\alpha}_s}{N} \frac{\Gamma(1-\epsilon)}{(-\epsilon)\pi^{-\epsilon}} \frac{(\mu^2 {\bf x}_T^2)^{\epsilon}}{n+1}J_n(\epsilon)\times \tilde{\cal C}_{n}(N,{\bf x}_T,\mu), \\
&& J_n(\epsilon)=n+1-\frac{n\Gamma(1-\epsilon)\Gamma(1-\epsilon(n+1))\Gamma(1+\epsilon)\Gamma(1+\epsilon(n-1))}{\Gamma(1-n\epsilon)\Gamma(1-2\epsilon)\Gamma(1+n\epsilon)}.
\end{eqnarray}

The advantage of working in ${\bf x}_T$-space is, that inverse Fourier transform of the product of evolution factor and ${\bf x}_T$-space coefficient function will not produce any additional divergences, so all collinear divergences are contained in the evolution factor in ${\bf x}_T$-space and can be subtracted from it. The renormalization factor which subtracts collinear divergences from the hard process can be defined in $N$-space as (see Eq.~(2.28) in Ref.~\cite{Catani:1994sq}):
\[
Z_{\rm coll.}(\epsilon,N)=\exp\left[ \frac{1}{\epsilon} \sum\limits_{k=1}^\infty \frac{(\hat{\alpha}_s S_\epsilon (\mu^2/\mu_F^2)^\epsilon)^k}{k}\gamma_k(N) \right],
\]
where $S_\epsilon=\exp\left[\epsilon (\log 4\pi - \gamma_E) \right]$ is the usual factor defining the $\overline{MS}$-scheme, $\mu_F$ is the factorization scale and $\gamma_k(N)$ are the coefficients of expansion of DGLAP anomalous dimension $\gamma_{gg}(N,\alpha_s)$ in powers of $\hat{\alpha}_s$. In agreement with known results~\cite{Jaroszewicz:1982gr, Catani:1994sq}, we find that the following series of coefficients leads to subtraction of collinear divergences from $\tilde{\cal C}(N,{\bf x}_T,\mu)$ up to $O(\hat{\alpha}_s^{9})$:
\begin{eqnarray*}
\gamma_1=\frac{1}{N},\ \gamma_2=\gamma_3=\gamma_5=0,\ \gamma_4=\frac{2\zeta(3)}{N^4},\\ \gamma_6=\frac{2\zeta(5)}{N^6},\ \gamma_7=\frac{12\zeta^2(3)}{N^7},\ \gamma_8=\frac{2\zeta(7)}{N^8},\ \gamma_9=\frac{32\zeta(3)\zeta(5)}{N^9} .
\end{eqnarray*}

We have checked up to $O(\hat{\alpha}_s^{9})$, that the finite part of the coefficient function can be represented as:
\begin{eqnarray*}
\tilde{\cal C}_{\rm coll. ren.}(N,{\bf x}_T,\mu)&=&\exp\left[ -\frac{\hat{\alpha}_s}{N} \log(\mu_F^2 \bar{\bf x}_T^2) \right]\\ &\times&\left\{ 1 + \frac{\hat{\alpha}_s^3}{N^3}2\zeta(3)-\frac{\hat{\alpha}^4_s}{N^4} \left(2\zeta(3) \log(\mu_F^2 \bar{\bf x}_T^2) + \frac{\pi^4}{120}  \right) + O(\hat{\alpha}_s^5) \right\},
\end{eqnarray*}
where $\bar{\bf x}_T={\bf x}_T e^{\gamma_E}/2$, the non-cancellation of $\gamma_E$ and $1/2$ is a consequence of working in ${\bf x}_T$-space. The exponential factor in last equation resums double-logarithms of the form $\hat{\alpha}_s \log({\bf x}_T^2\mu_F^2)\log(1/x)$ and corrections to it are at best -- single-logarithmic and start at $O(\alpha_s^3)$. Therefore, the double-logarithmic approximation is the basic approximation for UPDF. Converting the exponential factor back to $x$-space one obtains:
\begin{equation}
\tilde{\cal C}_{\rm DL-pert.}(x,{\bf x}_T,\mu)=\delta(1-x)-\sqrt{\frac{\hat{\alpha}_s \log(\mu_F^2 \bar{\bf x}_T^2)}{\log x}} I_1\left(2\sqrt{\hat{\alpha}_s\log(\mu_F^2 \bar{\bf x}_T^2)\log(x)} \right), \label{eq:C-DL-pert}
\end{equation}
where $I_1(x)$ is the Bessel function of the first kind. Before Fourier-transforming this expression numerically back to ${\bf q}_T$-space, we multiply it by the non-perturbative shape-function, which we take in a Gaussian form, suppressing large values of ${\bf x}_T$:
\[
F_{\rm NP}({\bf x}_T)=\exp\left[-\Lambda^2 {\bf x}_T^2 \right],
\]
where parameter $\Lambda$, equal to 1 GeV in our numerical calculations, characterizes the spread of ``intrinsic'' transverse-momentum of a gluon in a proton. Finally, to obtain the UPDF we take a Mellin convolution of the evolution factor with the collinear PDF as in Eq.~(\ref{eq:UPDF-def}). In the numerical calculations of the present paper we have used the NLO set of CTEQ-14 PDFs~\cite{Dulat:2015mca} as a collinear input and the NLO running of $\alpha_s$ corresponding to this PDF set with $\alpha_s(M_Z)=0.106$, as provided by LHAPDF library~\cite{Buckley:2014ana}. 

\begin{figure}
\begin{center}
\includegraphics[width=0.45\textwidth]{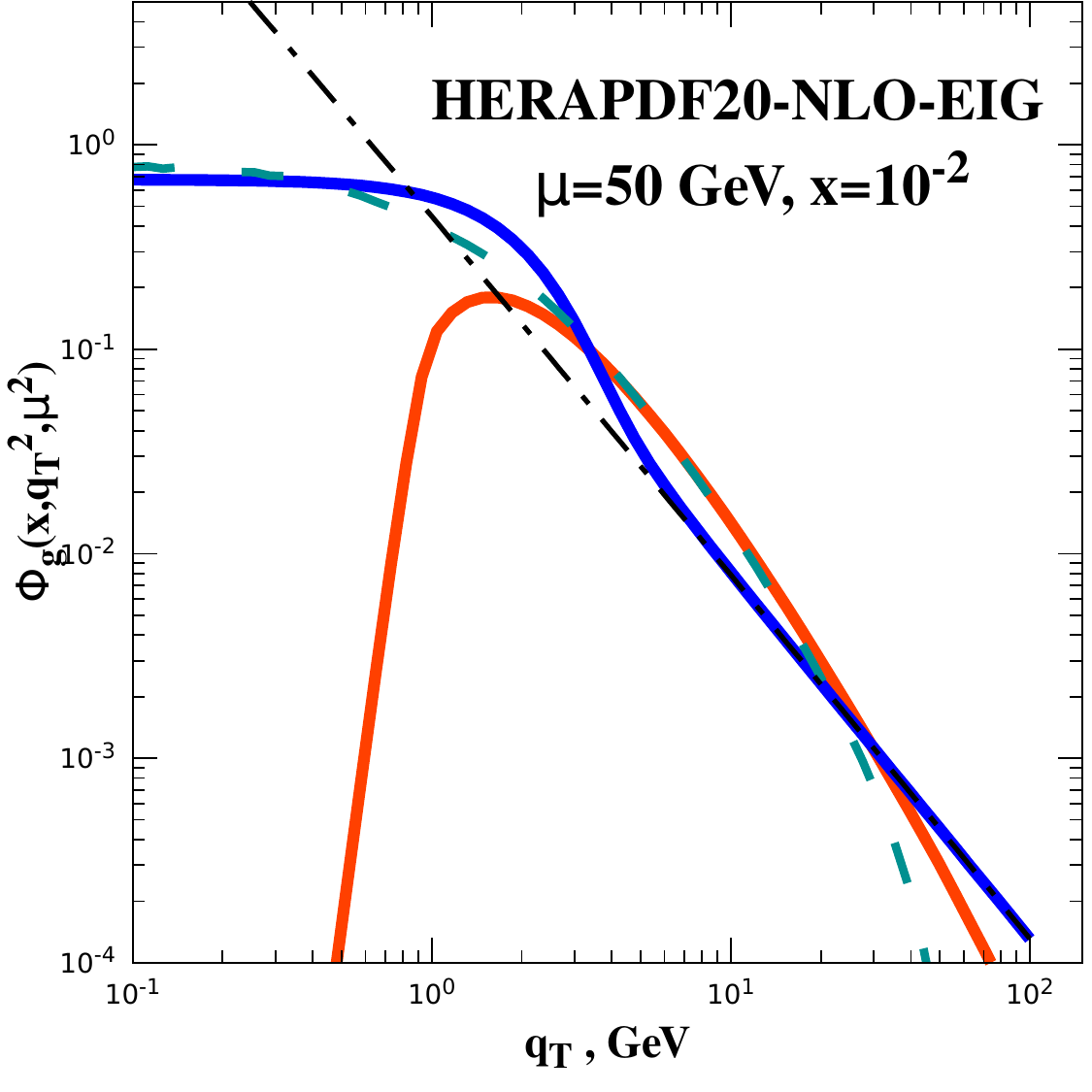}
\includegraphics[width=0.45\textwidth]{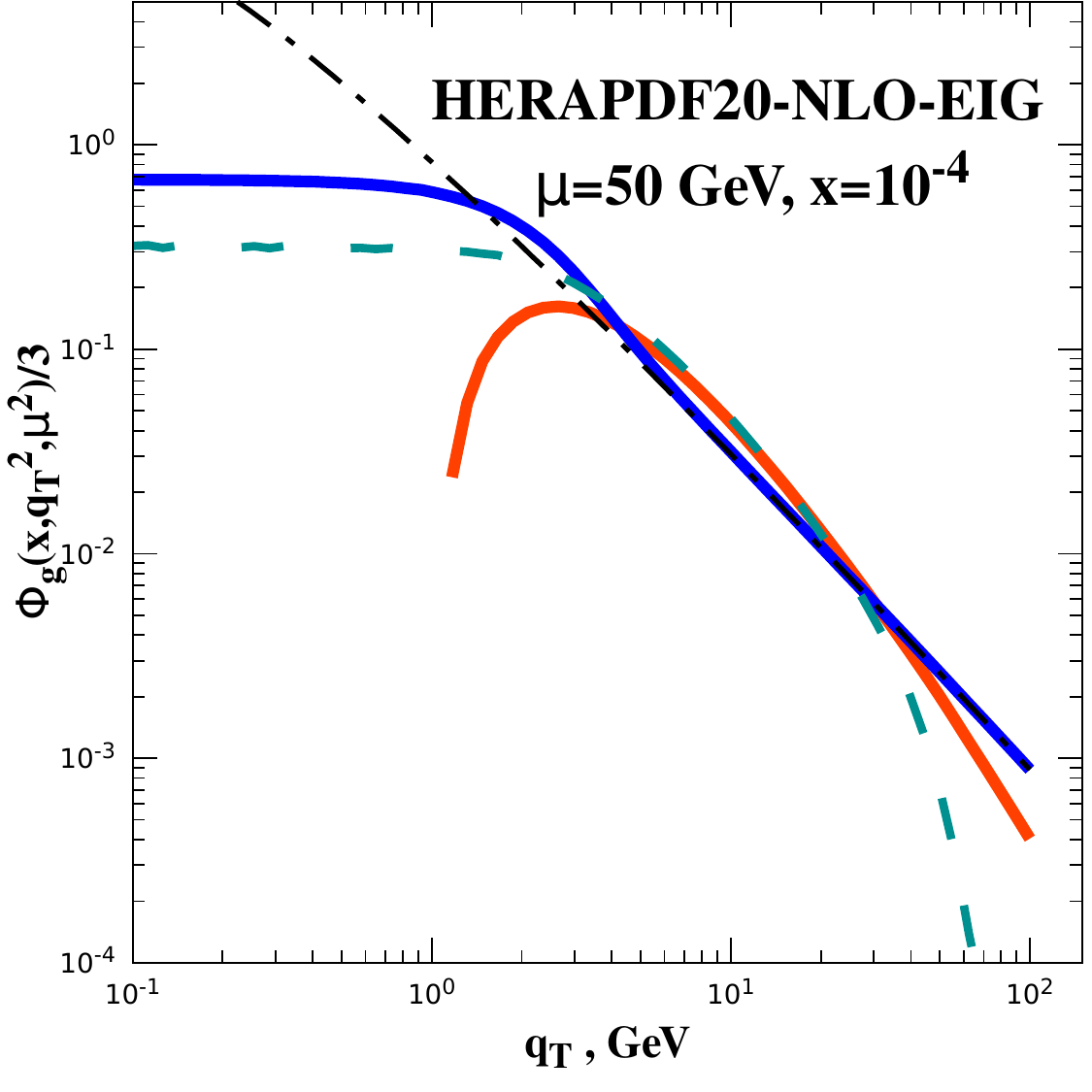}
\end{center}
\caption{Comparison of ${\bf q}_T$-dependence of Doubly-logarithmic UPDF (solid blue line) proposed in the present paper with several widely-used phenomenological UPDFs: solid orange line -- KMRW~\protect\cite{Kimber:2001sc, Watt:2003mx, Watt:2003vf} formula with exact normalization~\protect\cite{NS_DIS1}, dashed green line -- Parton-Branching method~\protect\cite{Martinez:2018jxt,Hautmann:TMDlib}, black dash-dotted line -- Collins-Ellis-Bl\"umlein doubly-logarithmic approximation~\protect\cite{Collins:1991ty, Blumlein:1995eu}, all with the parameters $\mu_F=\mu_R=50$ GeV and $x=10^{-2}$ (left panel) and $10^{-4}$ (right panel, UPDFs are multiplied by $1/3$). All UPDFs are obtained from HERAPDF20-NLO-EIG PDF set~\protect\cite{Abramowicz:2015mha,Buckley:2014ana} as a collinear input. \label{fig:UPDFs}}
\end{figure}

  In the Fig.~\ref{fig:UPDFs} we compare Doubly-Logarithmic UPDF with several other phenomenological UPDFs known in the literature. All UPDFs in this figure are obtained from HERAPDF-20-NLO-EIG PDF set~\cite{Abramowicz:2015mha,Buckley:2014ana} as a collinear input. In general, one observes that all distributions are broadening in ${\bf q}_T^2$ with decreasing $x$. There is a serious disagreement between different approaches at small values of $|{\bf q}_T|\leq 1$ GeV, since no dedicated fits where performed there. At moderate 1 GeV $<|{\bf q}_T|<\mu$ all approaches basically agree in shape, which is a manifestation of universality of doubly-logarithmic approximation. For $|{\bf q}_T|>\mu$ Doubly-logarithmic and Collins-Ellis-Bl\"umlein UPDFs continue the same trend, while KMRW and PB UPDFs demonstrate more physical softer behavior. In general, the latter UPDFs tend to agree with data better, while Collins-Ellis-Bl\"umlein tends to over-estimate absolute values of cross-sections. Agreement in shape and normalization of high-${\bf q}_T^2$ tail between Collins-Ellis-Bl\"umlein UPDF and our doubly-logarithmic approach is not surprising, since both approaches resum the same tower of double logarithms $\log(\mu^2/{\bf q}_T^2)\log(1/x)$, but  Collins-Ellis-Bl\"umlein approach is set-up directly in ${\bf q}_T$-space and is less systematic from the point of view of keeping correct $\overline{MS}$-scheme definition. Also, Collins-Ellis-Bl\"umlein-approach does not contain any non-perturbative shape-function, which explains different small-${\bf q}_T$ behavior.

\section*{Appendix B: Taking iterations of MRK-evolution kernel}
\hypertarget{sec:AppendixB}{}

  In this appendix we will study the divergence structure of the Eq.~(\ref{eq:C-evol-MRK}) by computing it's iterations. Taking one iteration of a virtual part with initial condition (\ref{eq:C1}) one obtains:
\begin{eqnarray}
{\cal C}_2^{\rm (V)}(x,{\bf q}_T,\mu_Y)&=& \frac{\alpha_s C_A}{\pi} \frac{2\omega_g({\bf q}_T^2)}{{\bf q}_T^2} \int\limits_x^1\frac{dz}{z(1-z)} \nonumber \\ &\times& \theta\left[\Delta\left(|{\bf q}_T|,\mu_Y\frac{x(1-z)}{z(z-x)}\right)-\frac{x}{z}\right] \theta(\Delta(|{\bf q}_T|,\mu_Y)-z) \nonumber \\
&=&\frac{\alpha_s C_A}{\pi} \frac{2\omega_g({\bf q}_T^2)}{{\bf q}_T^2} \log\left[ \frac{\mu_Y}{|{\bf q}_T|} \frac{1-x}{x} \right] \theta(\Delta(|{\bf q}_T|,\mu_Y)-x).\label{eq:C2V}
\end{eqnarray}

  The logarithm in Eq. (\ref{eq:C2V}) is the difference between $Y_\mu=\log(\mu_Y/q_1^-)$ and rapidity of the first real emission in the cascade: $y_1=\log(|{\bf q}_T|x/(q_1^-(1-x)))$, as one would expect from Eq.~(\ref{eq:BFKL-Y}). Iterating the virtual part of rapidity-space evolution equation (\ref{eq:BFKL-Y}) one finds that 
\[
{\cal C}_n^{\rm (V)}(Y_\mu,{\bf q}_T)=\frac{\alpha_s C_A}{\pi {\bf q}_T^2} \frac{\left[ 2\omega_g({\bf q}_T^2) (Y_\mu-y_1) \right]^{n-1}}{(n-1)!} \theta(Y_\mu-y_1),
\]  
so that the Regge-trajectory contribution exponentiates as $\exp\left[2\omega_g({\bf q}_T^2) (Y_\mu-y_1) \right]$. The $x$-space version of evolution equation should follow the same pattern, as it is easy to check e.g. calculating the next iteration:
\begin{eqnarray*}
{\cal C}_3^{\rm (V)}(x,{\bf q}_T,\mu_Y)&=& \frac{\alpha_s C_A}{\pi} \frac{\left[2\omega_g({\bf q}_T^2)\right]^2}{{\bf q}_T^2} \int\limits_x^1\frac{dz}{z(1-z)} \log\left[ \frac{\mu_Y}{|{\bf q}_T|} \frac{1-z}{z} \right] \nonumber \\ &\times& \theta\left[\Delta\left(|{\bf q}_T|,\mu_Y\frac{x(1-z)}{z(z-x)}\right)-\frac{x}{z}\right] \theta(\Delta(|{\bf q}_T|,\mu_Y)-z) \nonumber \\
&=&\frac{\alpha_s C_A}{\pi} \frac{\left[2\omega_g({\bf q}_T^2)\right]^2}{{\bf q}_T^2} \frac{1}{2}\log^2\left[ \frac{\mu_Y}{|{\bf q}_T|} \frac{1-x}{x} \right] \theta(\Delta(|{\bf q}_T|,\mu_Y)-x).
\end{eqnarray*}

To see an example of the cancellation of infra-red divergences let's consider the second iteration of the real-emission kernel of (\ref{eq:C-evol-MRK}):
\begin{eqnarray}
&&{\cal C}_2^{\rm (R)}(x,{\bf q}_T,\mu_Y)=\left(\frac{\alpha_sC_A}{\pi}\right)^2 \int\frac{d^{D-2}{\bf k}_T}{\pi(2\pi)^{-2\epsilon}} \frac{1}{{\bf k}_T^2({\bf q}_T+{\bf k}_T)^2} \times \nonumber \\
&&\times\int\limits_x^1\frac{dz}{z(1-z)}\theta\left[ \Delta\left({\bf k}_T+{\bf q}_T,\frac{|{\bf k}_T|}{1-z} \right) - \frac{x}{z}\right] \theta(\Delta(|{\bf q}_T|,\mu_Y)-z) = \nonumber \\
&&= \left(\frac{\alpha_sC_A}{\pi}\right)^2 \int\frac{d^{D-2}{\bf k}_T}{\pi(2\pi)^{-2\epsilon}} \frac{1}{{\bf k}_T^2({\bf q}_T+{\bf k}_T)^2} \nonumber \\
&& \times \log\left[ \frac{\mu_Y (1-x)}{x(|{\bf k}_T|+|{\bf q}_T+{\bf k}_T|)} \right]\theta\left[ \frac{1-x}{x} - \frac{|{\bf k}_T|+|{\bf q}_T+{\bf k}_T|}{\mu_Y} \right]. \label{eq:C2R}
\end{eqnarray}

  Two momentum regions generate divergences in Eq.~(\ref{eq:C2R}): the region $|{\bf k}_T|\ll |{\bf q}_T|$ where the second emission is soft, while the transverse momentum is generated by the first emission, and $|{\bf k}_T+{\bf q}_T|\ll |{\bf q}_T|$, where the first emission is collinear, while the second emission generates transverse momentum. The $1/\epsilon$ pole of the infra-red divergent contribution from the first region:
\[
 \left(\frac{\alpha_sC_A}{\pi}\right)^2 \frac{1}{{\bf q}_T^2} \log\left[ \frac{\mu_Y}{|{\bf q}_T|} \frac{1-x}{x} \right] \int\limits^\Lambda \frac{d^{D-2}{\bf k}_T}{\pi {\bf k}_T^2},
\]
is cancelled by the virtual contribution (\ref{eq:C2V}), while the divergence from {\it collinear} region -- remains. It is natural to expect the collinear divergence to appear, since we have two real emissions and one of them can take whole transverse momentum ${\bf q}_T$, while another generates collinear divergence which should be absorbed into collinear PDF.

\bibliographystyle{JHEP}
\bibliography{mybibfile}

\end{document}